\documentclass[12pt]{article}
\usepackage[breaklinks=true]{hyperref}
\usepackage{hyperref} 
\hypersetup{colorlinks=true, linkcolor=black, citecolor=black, urlcolor=blue!80!black} 
\usepackage{amsfonts,amsmath,amsxtra,amsthm}
\usepackage{amssymb}
\usepackage{bm}
\usepackage[utf8]{inputenc}
\usepackage{lscape}
\usepackage[normalem]{ulem}
\usepackage[russian,english]{babel}
\usepackage{xcolor}
\usepackage{tikz, pgfplots}
\usepackage{tcolorbox}
\usepackage{mathrsfs}
\usepackage{cancel}

\numberwithin{equation}{section}

\usepackage{tocloft} 
\renewcommand{\epsilon}{\varepsilon}
\let\ggr\gg

\def\a{\alpha}

\def\b{\beta}
\def\bg{\boldsymbol{\gamma}}

\def\beq{\begin{equation}}
	\def\eeq{\end{equation}}
\def\beqq{\begin{equation*}}
	\def\eeqq{\end{equation*}}

\def\bs{\begin{split}}
	\def\es{\end{split}}
\def\bg{\boldsymbol{\gamma}}
\def\bl{{\boldsymbol{\lambda}}}
\def\bbg{\underline{\bm{\gamma}}}

\def\bbl{\underline{\boldsymbol{\lambda}}}

\def\bnu{\boldsymbol{\nu}}
\def\bbnu{\underline{\bnu}}
\def\bo{\boldsymbol{\omega}}
\def\brho{\boldsymbol{\rho}}

\def\bx{\boldsymbol{x}}
\def\by{\boldsymbol{y}}
\def\bz{\boldsymbol{z}}
\def\bk{\boldsymbol{k}}

\def\bbx{\underline{\boldsymbol{x}}}
\def\bby{\underline{\boldsymbol{y}}}

\def\bbz{\underline{\boldsymbol{z}}}

\def\const{{2\pi\imath}}

\def\dd{\hat{d}}
\def\ddelta{\hat{\Delta}}

\def\Im{\operatorname{Im}}
\def\cg{\mathit{g}}
\def\g{\gamma}
\def\gg{{\hat{g}^\ast}}
\def\gl{\mathfrak{gl}}

\def\H{\operatorname{H}}
\def\i{\imath}

\def\j{\hat{j}}

\def\k{\bar{k}}

\def\l{\lambda}

\def\mg{m_g}

\def\mm{\hat{\mu}}
\def\ppi{\sigma}
\def\rd{d^R}
\def\rg{\mathrm{g}}
\def\rm{w^R}
\def\rrm{\hat{w}^R}

\def\o{\omega}

\def\R{\mathbb{R}}
\def\Re{\mathrm{Re}\,}
\def\Res{\operatorname{Res}}

\def\ve{\varepsilon}
\def\vf{\varphi}

\def\Z{\mathbb{Z}}

	\newtheorem{lemma}{Lemma}[section]
	\newtheorem{proposition}{Proposition}[section]
	\newtheorem*{proposition*}{Proposition}
	\newtheorem{corollary}{Corollary}[section]
	\newtheorem*{theorem*}{Theorem}
	\newtheorem{theorem}{Theorem}[section]
	\newtheorem{remark}{Remark}[section]
	\newcommand{\rf}[1]{(\ref{#1})}

	\def\rPsi{\Psi^{R}} \def\cPsi{\Psi^{}} \def\rrPsi{\hat{\Psi}^{R}} \def\ccPsi{\hat{\Psi}^{}}
	\def\rQ{Q^{R}}\def\rrQ{\hat{Q}^{R}} \def\ccQ{\hat{Q}}
	\def\rM{M^{R}} \def\ccM{\hat{M}}
		\def\rL{\Lambda^{R}}\def\rrL{\hat{\Lambda}^{R}}
		 \def\cL{\Lambda}\def\ccL{\hat{\Lambda}}
		\def\rK{K^{R}}\def\rrK{\hat{K}^{R}} \def\cK{K}\def\ccK{\hat{K}}
		\def\rmu{\mu^{R}} \def\cmu{\mu}\def\ccmu{\hat{\mu}}
		\def\ccm{\hat{w}}

	\renewcommand{\iota}{\imath}
        \DeclareMathOperator{\sign}{sign}
	
\begin{document}
	
	\begin{center}
			{\bf \large Calogero--Sutherland hyperbolic system and Heckman--Opdam $\gl_n$ hypergeometric function \\[4pt] }
			
			\vspace{0.4cm}
			
			{N. Belousov$^{\dagger}$, L. Cherepanov$^{\circ}$, S. Derkachov$^{\dagger\times}$, S. Khoroshkin$^{\dagger\ast\star}$}
			
			\vspace{0.4cm}
			
			{\small \it
			$^\dagger$Beijing Institute of Mathematical Sciences and Applications, \\
				Huairou district, Beijing, 101408, China \vspace{0.2cm}\\
				$^\times$Steklov Mathematical Institute, Fontanka 27, \\
				St. Petersburg, 191023, Russia \vspace{0.2cm}\\
			$^\ast$Department of Mathematics, Technion,
				Haifa, 32000, Israel \vspace{0.2cm}\\
			$^\circ$National Research University Higher School of Economics, \\
		Myasnitskaya 20, Moscow, 101000, Russia \vspace{0.2cm}	\\
				$^\star$Skolkovo Institute of Science and Technology,\\Skolkovo, 121205, Russia }
		\end{center}
		
		\begin{abstract} 
			We prove equivalence of two integral representations for the wave functions of hyperbolic Calogero--Sutherland system. For this we study two families of Baxter operators related to hyperbolic Calogero--Sutherland and rational Ruijsenaars models; the first one as a limit from hyperbolic Ruijsenaars system, while the second one independently. Besides, computing asymptotics of integral representations and also  the value at zero point, we identify them with renormalized Heckman--Opdam~$\gl_n$ hypergeometric function. 
		\end{abstract}
		\tableofcontents
		
		\section{Introduction}
		Recently, significant progress has been made in the study of the Ruijsenaars quantum hyperbolic model introduced in \cite{Ru1}. In \cite{Ru3} S. Ruijsenaars found the so called ``kernel function'' playing the role of reproducing kernel for the model. By use of the kernel function in~\cite{HR1} M. Halln\"as and S. Ruijsenaars constructed integral representation for wave functions of the model. Besides, in a series of papers \cite{BDKK1,BDKK2,BDKK3,BDKK5} the technique of integral Baxter operators for the Ruijsenaars model was developed, which helped to establish bispectral and coupling constant dualities together with the orthogonality and completeness relations. 
		
		It is natural to expect that similar statements take place for nonrelativistic limit of Ruijsenaars hyperbolic model, the hyperbolic Calogero--Sutherland model,  governed by the Hamiltonian
		\begin{align} \label{suth-ham}
			\mathcal{H} = - \sum_{j = 1}^n \partial_{x_j}^2 + \sum_{1 \leq j \not= k \leq n} \frac{ \pi^2 g (g - 1)}{\sh^2 \pi(x_j - x_k)}. 
		\end{align}
		Moreover, almost all ingredients for that are known. The following Euler type integral representation of wave functions was found by M. Halln\"as and S. Ruijsenaars in \cite{HR0}
		\begin{multline} \label{Psi-E}
			\Psi_{\bl_n}(\bx_n) = \prod_{k = 1}^{n - 1} \frac{(2\pi \Gamma(g))^k}{k!} \!\! \! \! \! \!  \int\limits_{\mathbb{R}^{n(n - 1)/2}}\!\!\! \exp \biggl( 2\pi \imath \sum_{k = 1}^n \lambda_k \biggl( \sum_{i = 1}^{k} y_{ki} - \sum_{i = 1}^{k - 1} y_{k - 1, i} \biggr) \biggr) \\[6pt]
			\times  \prod_{k = 1}^{n - 1} \frac{\prod_{1 \leq i < j \leq k} \, [4 \sh^2 \pi (y_{ki} - y_{kj})]^g}{ \prod_{i = 1}^{k + 1} \prod_{j = 1}^{k} \, [2\ch \pi (y_{k+ 1, i} - y_{kj})]^g }  \, \prod_{k = 1}^{n - 1} \prod_{i = 1}^k dy_{ki},
		\end{multline}
		where we denote $\bx_n = (x_1, \dots, x_n)$ and identify $y_{ni} \equiv x_i$. The related Baxter operators were studied by M.~Halln\"as~\cite{H}. The different expression by means of Mellin--Barnes integrals 
		\begin{multline} \label{Psi-MB}
			\hat{\Psi}_{\bl_n}(\bx_n) = \prod_{k = 1}^{n - 1} \frac{(2\pi \Gamma(g))^{-k}}{k!} \!\!\!\!\!\! \int\limits_{\mathbb{R}^{n(n - 1)/2}}\!\!\!\!\! \exp \biggl( 2\pi \imath \sum_{k = 1}^n x_k \biggl( \sum_{i = 1}^{k} \gamma_{ki} - \sum_{i = 1}^{k - 1} \gamma_{k - 1, i} \biggr) \biggr) \\[6pt]
			\times  \prod_{k = 1}^{n - 1} \frac{ \prod\limits_{i = 1}^{k + 1} \prod\limits_{j = 1}^{k} \, \Gamma \bigl( \imath \gamma_{k + 1, i} - \imath  \gamma_{kj} + \frac{g}{2} \bigr) \, \Gamma \bigl( \imath \gamma_{k j} - \imath  \gamma_{k + 1, i} + \frac{g}{2} \bigr) }{\prod_{1 \leq i \not= j \leq k} \, \Gamma(\imath \gamma_{k i} - \imath \gamma_{k j}) \Gamma(\imath \gamma_{kj} - \imath \gamma_{ki} + g)} \, \prod_{k = 1}^{n - 1} \prod_{i = 1}^k d\gamma_{ki},
		\end{multline}
		where we identify $\gamma_{ni} \equiv \lambda_i$, was found in \cite{KK1}.
		Besides, there is a well developed theory of Heckman--Opdam hypergeometric functions \cite{HO,O}, which analyses wave functions through their series representation. However, the precise correspondence between different parts of the story was not rigorously established, except for the case of two particles \cite{BDKK6}, and the goal of this paper is to fill this gap.
		 
		First, in Section~\ref{sec:hyp-ruij} we collect known results for relativistic hyperbolic Ruijsenaars model needed in the present paper. Then, in Section~\ref{sec:suth} we study its limit to nonrelativistic Calogero--Sutherland model. In this limit basic ingredients of all integrals --- the kernel function and the measure function --- tend pointwise to their nonrelativistic counterparts.  However, we cannot declare the uniform limit of the measure function, needed for establishing limits of integrals. Instead we prove the uniform bound for the measure function and thus prove the basic integral identity which is behind the commutativity of Baxter operators. Then their theory goes in parallel to that of Baxter operators in relativistic model. In this way we reprove the results of M. Halln\"as \cite{H}.
		  
		  The hyperbolic Ruijsenaars model is self-dual in a sense that corresponding wave functions diagonalize the same difference operators with respect to coordinates and spectral parameters~\cite{BDKK2}. In the case of hyperbolic Calogero--Sutherland system the dual model on the space of spectral parameters is governed by rational Ruijsenaars--Macdonald operators,  which we consider in Section~\ref{sec:rat-ruij}. The corresponding limiting procedure from hyperbolic model is different. Again, we have a pointwise limit of kernel and  measure functions,  but this is not a uniform limit, which can be used in integrals. Therefore here we develop the theory of Baxter operators and their properties independently. In particular, we give a direct proof of commutativity of dual Baxter operators, which is a simplified version of the hyperbolic derivation from \cite{BDKK1}. Besides, using the Mellin--Barnes integral formula for the wave function~\eqref{Psi-MB} we describe the spectrum of dual Baxter operators. 
		  
		  The equivalence between two integral representations of the wave functions~\eqref{Psi-E},~\eqref{Psi-MB}
		 \begin{align}
		 	\Psi_{\bl_n}(\bx_n) = \hat{\Psi}_{\bl_n}(\bx_n),
		 \end{align}
		 or bispectral duality, is proved in Section~\ref{sec:as-dual}. The proof is based on the diagonalization properties of two families of Baxter operators, acting in the spaces of spatial and spectral variables. Besides, it relies on the commutativity of these families with each other, which follows from the convergence of certain integrals and requires certain bounds on the wave functions. Here we follow the lines of the paper \cite{HR3} and obtain the necessary bounds using Mellin--Barnes representation. In the rational limit the bounds are slightly more delicate because of power factors in the asymptotics of the gamma function.
		  
		  
		 In Appendix~\ref{AppD} we derive asymptotics of the considered wave function 
		 \begin{align}
		 	\begin{aligned}
		 		\Psi_{\bl_n}(\bx_n) & \sim e^{ \pi  g \sum_{i = 1}^n x_i (2i - n - 1)} \\[6pt] 
		 		 & \times \sum_{\sigma\in S_n}e^{2\pi \imath \sum_{i = 1}^n x_i\l_{\sigma(i)}}\prod_{{1\leq i<j \leq n}} \Gamma(\imath \l_{\sigma(i)}- \imath\l_{\sigma(j)}) \, \Gamma(\imath \l_{\sigma(j)}- \imath\l_{\sigma(i)} + g)
		 	\end{aligned}
		 \end{align} 
		 in Weyl chamber of spatial parameters $x_1 \ggr x_2 \ggr \ldots \ggr x_n$.
		 Comparing it with the known asymptotics of $\gl_n$ Heckman--Opdam hypergeometric function $F_{\bl_n}(\bx_n)$,  normalized as $F_{\bl_n}(0, ..., 0) = 1$, in Section~\ref{sec:HO} we show,   using the uniqueness property of  Heckman--Opdam hypergeometric function~\cite{HO, O},  that  
		 \begin{align} \label{00} 
		 	F_{\bl_n}(\bx_n)= { \prod_{k=1}^n \frac{\Gamma(kg)}{\Gamma(g)}}\, \prod_{1\leq i\not=j \leq n}\Gamma^{-1}(\imath(\l_i-\l_j)+g) \, {\Psi}_{\bl_n}(\bx_n).
		 \end{align} 
		 As a corollary, we state two dual integral representations for the Heck\-man--Opdam hypergeometric function, as well as precise form of its Harish-Chandra series, which we obtain by residue evaluation of Mellin--Barnes integrals.
		 
		  In Appendix C we perform another check of the identification \rf{00} calculating the wave function at zero point $x_1 = \ldots = x_n = 0$. 
		 
		In conclusion, we note that a third integral representation of the $\mathfrak{gl}_n$ Heckman--Opdam hypergeometric function is given in \cite{BG}.  The related Baxter operators are considered in~\cite{KMS} in the context of trigonometric Calogero--Sutherland model.

		\section{Ruijsenaars hyperbolic system} \label{sec:hyp-ruij}
	The Ruijsenaars hyperbolic system is described by a set of commuting Ruijsenaars--Macdonald difference operators
	\begin{equation}
		\label{r1}
		\rM_r(\bx_n) = \sum_{\substack{I\subset[n] \\ |I|=r}}
		\prod_{\substack{i\in I \\ j\notin I}}
		\frac{\sh\frac{\pi}{\o_2}\left(x_i-x_j-\imath\rg\right)}
		{\sh\frac{\pi}{\o_2}\left(x_i-x_j\right)}
		\cdot T^{-\imath\o_1}_{I,x},
	\end{equation}
which act on functions of $n$ complex variables
 \begin{equation*}
	\bm{x}_n = (x_1, \dots,x_n)
\end{equation*}
analytical in a proper vicinity of $\R^n$. Here $T^{a}_{x_i}$ is the shift operator
\begin{equation}
	T^{a}_{x_i}=e^{a\partial_{x_i}}, \qquad \left( T^{a}_{x_i} \, f \right)(x_1,\ldots,x_n) = f(x_1,\ldots,x_i+a,\ldots,x_n)
\end{equation}
and  for any subset $I\subset[n] = \{1, \dots, n\}$
\begin{equation}
	T^{a}_{I,x}=\prod_{i \in I} T_{x_i}^a.
\end{equation}

The system is parametrized by three constants: periods $\bm{\omega}=(\omega_1, \omega_2)$ and coupling constant $g$, subject to conditions
\beq\label{r2} \Re \o_1 > 0, \qquad \Re \o_2 > 0,\qquad 0< \Re \rg<\Re \o_1+\Re \o_2  \eeq
and also
\beq\label{r3}  0 < \Re\frac{ \rg}{\o_1\o_2} < \Re \frac{\o_1 + \o_2}{\o_1 \o_2}.  \eeq
The operators $\rM_r$ are symmetric with respect to the bilinear form
\beq \label{r4}\left(f(\bx_n),\, g(\bx_n)\right)_{\rmu}=\int_{\R^n} f(\bx_n)g(-\bx_n)\rmu(\bx_n) d\bx_n. \eeq
Here the measure function $\rmu(\bx_n)$ is defined as follows.
First, define the weight function \beq\label{5a}\rm(x)=S_2(\imath x) S_2^{-1}(\imath x+\rg)=S_2(\imath x)S_2(-\imath x+\rg^\ast), \eeq
where $S_2(z)=S_2(z|\bo)$ is the double sine function, see Appendix~\ref{AppA}, and
\beq\label{r7} \rg^\ast=\o_1+\o_2-\rg. \eeq
Besides, denote
\beq\label{r7a} \rmu(x) := \rmu(x|\o_1, \o_2, \rg) = \rm(x)\rm(-x).
\eeq
Finally, we set
\beq\label{r5}\begin{split}
& \rm(\bx_n)=\prod_{1\leq i<j \leq n} \rm(x_i-x_j),\\
& \rmu(\bx_n)=\rm(\bx_n)\rm(-\bx_n)=\prod_{\substack{i,j=1 \\ i\not=j}}^n\rm(x_i-x_j)=\prod_{1\leq i<j \leq n} \rmu(x_i-x_j).\end{split} \eeq

To describe the wave functions and Baxter operators we also need the following function of a complex variable
\beq\label{r6} \rK(x):= \rK(x| \o_1, \o_2, \rg) = S_2^{-1}\Bigl(\imath x +\frac{\rg^\ast}{2} \Bigr)S_2^{-1}\Bigl(-\imath x+\frac{\rg^\ast}{2} \Bigr).\eeq
 The Ruijsenaars kernel function is defined as  the product
\begin{equation}\label{r8}
	\rK(\bx_n,\by_m)=\prod_{i=1}^n \prod_{j = 1}^m \rK(x_i-y_j).
\end{equation}
One of the key objects for the analysis of the system is a family of Baxter operators $\rQ_{n}(\lambda)$ parameterized by $\lambda \in \mathbb{C}$. These are integral operators
\begin{equation}\label{r9}
	\left( \rQ_{n}(\lambda) f\right) (\bm{x}_n) =  \rd_n(\rg) \, \int_{\mathbb{R}^n} d\bm{y}_n \, \rQ(\bm{x}_n, \bm{y}_n; \lambda) f(\bm{y}_n)
\end{equation}
with the kernel
\beq\label{r10} \rQ(\bx_n,\by_{n};\l)= e^{\const \l(\bbx_n-\bby_n)}
\rK(\bx_n,\by_{n})\rmu  (\by_{n}),
\eeq
and normalizing constant
\begin{equation}\label{dconst}
	\rd_n(\rg) = \frac{1}{n!} \left[ \sqrt{\omega_1 \omega_2} S_2(\rg) \right]^{-n}.
\end{equation}
Here and below we use the notation
$$ \bbx_n= \sum_{j=1}^nx_j. $$
Baxter operators commute between themselves as well as with Hamiltonians $\rM_r$ \cite{BDKK1}:
\begin{align}\label{r12}
	& \rQ_{n}(\lambda) \, \rQ_{n}(\rho) = \rQ_{n}(\rho) \, \rQ_{n}(\lambda),\\[6pt]
	\label{r13}
	& \rQ_{n}(\lambda) \, \rM_r(\bm{x}_n) = \rM_r(\bm{x}_n) \, \rQ_{n}(\lambda), \qquad r=1, \ldots, n.
\end{align}
Next, denote by $\rL_{n}(\l)$ the integral operator similar to the Baxter $Q$-operator
\beq\label{r14}\begin{split} \left(\rL_{n}(\l)f\right)(\bx_n)=\rd_{n - 1} \int_{\R^{n-1}}d\by_{n-1} \, \rL(\bx_n,\by_{n-1};\l) f(\by_{n-1})
\end{split}\eeq
with the kernel
\beq\label{r15} \rL(\bx_n,\by_{n-1};\l)= e^{\const \l(\bbx_n-\bby_{n-1})}
\rK(\bx_n,\by_{n-1})\rmu  (\by_{n-1}), 
\eeq
and constant $d^R_{n - 1}$ given by the formula \eqref{dconst}.
According to \cite{HR1} (see also~\cite{BDKK2}) the function
\beq \label{r16}\rPsi_{\bl_n}(\bx_n):=\rPsi_{\bl_n}(\bx_n; \rg|\bo)=\rL_{n}(\l_n) \, \rL_{n-1}(\l_{n-1})\cdots \rL_{2}(\l_2) \, e^{\const \l_1x_1}
\eeq
is a joint eigenfunction of Macdonald operators:
	\beq \label{r17} \rM_r(\bx_n)\,\rPsi_{\bl_n}(\bx_n)= e_r \bigl(e^{{2\pi \lambda_1}{\o_1}}, \dots, e^{{2\pi \lambda_n}{\o_1}}\bigr) \rPsi_{\bl_n}(\bx_n).
	\eeq
	Here $e_r(z_1,\ldots,z_n)$ is $r$-th elementary symmetric function,
	\beqq e_r(z_1,\ldots,z_n)=\sum_{1\leq i_1<i_2<\ldots< i_r\leq n}z_{i_1}\cdots z_{i_r}. \eeqq 
	Let us remark that due to symmetry of the double sine function $S_2(z|\o_1, \o_2) = S_2(z|\o_2, \o_1)$, the function $\rPsi_{\bl_n}(\bx_n)$ is symmetric with respect to periods. Hence, it also diagonalizes the Macdonald operators with interchanged periods $\o_1 \leftrightarrows \o_2$.
	
	 In what follows $\rL_{n}(\lambda)$ is called raising operator. These operators satisfy exchange relation
	\begin{align}
		\rL_n(\l) \rL_{n - 1}(\rho) =  \rL_{n }(\rho) \rL_{n - 1}(\l),
	\end{align}
	see~\cite{BDKK2}, which leads to the symmetry of wave functions~\eqref{r16} with respect to spectral variables.
	
To describe further properties of Baxter operators we need the following notation
	\beq \label{r18}\hat{\bo}=\frac{\bo}{\o_1\o_2}=\left(\frac{1}{\o_2},\frac{1}{\o_1}\right) ,
	\qquad \hat{\rg}=\frac{\rg^*}{\o_1\o_2}=\hat{\o}_1 + \hat{\o}_2 - \frac{\rg}{\o_1\o_2}.
	\eeq
	 Moreover, for any quantity $a$, which depends on parameters $\bo$ and $\rg$ we denote by $\hat{a}$ the same quantity with parameters $\bo$ and $\rg$ replaced by $\hat{\bo}$ and $\hat{\rg}$. For instance, 
	\beq\label{r19}\begin{split} 
		& \rrm(\l)=S_2(\imath \l|\hat{\bo}) S_2^{-1}(\imath \l+\hat{\rg}|\hat{\bo}), \\[6pt]
		& \rrK(\l) =S_2^{-1}\Bigl(\imath \l +\frac{\hat{\o}_1 + \hat{\o}_2 - \hat{\rg}}{2}\Big|\hat{\bo}\Bigr)S_2^{-1}\Bigl(-\imath \l+\frac{\hat{\o}_1 + \hat{\o}_2 - \hat{\rg}}{2}\Big|\hat{\bo}\Bigr).\end{split}\eeq
A degeneration of commutativity relation \rf{r12} gives the exchange relation
\begin{equation}\label{r20}
	\rQ_{n}(\lambda) \, \rL_{n}(\rho) = \rrK(\lambda - \rho) \; \rL_{n}(\rho) \, \rQ_{n-1}(\lambda),
\end{equation}
which in its turn implies that $\rPsi_{\bl_n}(\bx_n)$ diagonalizes Baxter operators
	\begin{equation}\label{r21}
\rQ_{n}(\lambda) \, \rPsi_{ \bm{\lambda}_n }(\bm{x}_n) = \prod_{j = 1}^n \rrK(\lambda-\lambda_j) \, \rPsi_{ \bm{\lambda}_n }(\bm{x}_n),
\end{equation}
the details are given in~\cite{BDKK2}.

 Using the notations \rf{r18} let us introduce dual Baxter and raising operators, acting in the spaces of spectral variables
  \begin{equation}\label{r22}\begin{aligned}
  		& \bigl( \rrQ_n(x) f\bigr) (\bm{\l}_n) = \rd_n(\hat{\rg}|\hat{\bo})\, \int_{\mathbb{R}^n} d\bm{\gamma}_n \, \rrQ(\bm{\l}_n, \bm{\gamma}_n; x) f(\bm{\gamma}_n),\\[7pt]
  		& \bigl(\rrL_{n}(x)f\bigr)(\bl_n) =\rd_{n - 1}(\hat{\rg}|\hat{\bo}) \int_{\R^{n-1}}d\bg_{n-1} \, \rrL(\bl_n,\bg_{n-1};x) f(\bg_{n-1})
  	\end{aligned}
  \end{equation}
  with the kernels
  \beq\label{r23}\begin{split} 
  	& \rrQ(\bl_n,\bg_{n};x)= e^{\const x (\bbl_n-\bbg_n)}	\rrK(\bl_n,\bg_{n})\hat{\mu}(\bg_{n}),\\[7pt]
  	& \rrL(\bl_n,\bg_{n-1};x) = e^{\const x (\bbl_n-\bbg_{n-1} )} \rrK(\bl_n,\bg_{n-1}) \hat{\mu}  (\bg_{n-1}).
  \end{split}
  \eeq
  Dual Baxter operators commute between themselves and obey the corresponding exchange relation with dual raising operators
  \begin{align}\label{r24}
  	& \rrQ_{n}(x) \, \rrQ_{n}(y) =  \rrQ_{n}(y) \, \rrQ_{n}(x),\\[6pt] \label{r25}
  	& \rrQ_{n}(x) \, \rrL_{n}(y) =  \rK(x - y) \; \rrL_{n}(y) \, \rrQ_{n-1}(x).
  \end{align}
The duality theorem \cite{BDKK2} states that the wave function \rf{r16} can be also constructed via dual raising operators
 \beq \rPsi_{\bl_n}(\bx_n)=\rrL_{n}(x_n) \, \rrL_{n-1}(x_{n-1})\cdots \rrL_{2}(x_2) \, e^{\const \l_1x_1},
\eeq
or in short,
\beq \label{r26} \rrPsi_{\bx_n}(\bl_n)= \rPsi_{\bl_n}(\bx_n). \eeq
Due to this property and also symmetry with respect to the periods $\o_1 \leftrightarrows \o_2$, wave functions not only diagonalize the operators~\eqref{r1}, but also
\begin{align} \label{MR-dual}
	\hat{M}^R_r:= \sum_{\substack{I\subset[n] \\ |I|=r}} \prod_{\substack{i\in I \\ j\notin I}} \frac{\sh \pi \o_2 \bigl(\l_i-\l_j-\imath \frac{\omega_1 + \omega_2 - \rg}{\omega_1 \omega_2} \bigr)}{\sh \pi \o_2(\l_i-\l_j)} \cdot T^{-\imath/\o_1}_{I,\l},
\end{align}
that is
\begin{align}
	\hat{M}^R_r \, \rPsi_{\bl_n}(\bx_n) = e_r \bigl( e^{2\pi x_1/\o_1}, \ldots, e^{2\pi x_n/\o_1} \bigr) \, \rPsi_{\bl_n}(\bx_n) .
\end{align}

\section{Calogero--Sutherland hyperbolic system} \label{sec:suth}
	The nonrelativistic counterparts of the kernel, weight and measure functions are
	\beq\label{s1}  \cK(x)=(2\ch \pi x)^{-\cg}, \qquad w(x)=|2\sh \pi x|^{\cg},\qquad \mu(x)=w^2(x)\eeq
	and, correspondingly,
	\beq\label{s2} \begin{split} \cK(\bx_n,\by_m)=\prod_{i = 1}^n \prod_{j = 1}^m \cK(x_i-y_j),\qquad & w(\bx_n)= \prod_{1\leq i<j \leq n} w (x_i-x_j),\\[3pt]  & \mu(\bx_n)= \prod_{1\leq i<j \leq n} \mu(x_i-x_j).\end{split}\eeq
	In these notations, the commuting Hamiltonians of (reduced) hyperbolic Calogero--Sutherland $n$-particle system can be written as follows \cite[eq. (11)]{HR1}:
	\beq\label{s3} \H_k=w^{-1}(\bx_n)\cdot{\mathcal{H}}_k\cdot w(\bx_n),
	\eeq
	where 
	\beq\begin{split}\label{s4} \mathcal{H}_k &=\frac{1}{(n-k)!}\sum\limits_{0\leq l\leq\left[\frac{k}{2}\right]}\frac{1}{2^l l!(k-2l)!}\\& \times
    \sum\limits_{\sigma\in S_n}s_g(x_{\sigma(1)}-x_{\sigma(2)}) \cdots s_g(x_{\sigma(2l-1)}-x_{\sigma(2l)}) \, p_{\sigma(2l+1)}\cdots p_{\sigma(k)},
	\end{split}\eeq
	\beq\label{s5}  s_g(x)=\frac{\pi^2 g(1-g)}{\sh^2\pi x},\qquad p_j=-\imath \partial_{x_j}. 
	\eeq 
	 Note that the Hamiltonian \eqref{suth-ham} is given by the combination $\mathcal{H}=\mathcal{H}_1^2 - 2\mathcal{H}_2$.
	
	All the results of this section hold under assumption on the coupling constant
	\begin{align}
		g > 0,
	\end{align}
	whereas starting from Section~\ref{sec:rat-ruij} we impose the stronger condition $g > 1$.  At the end this restriction can be relaxed after renormalization of the wave function and its identification with Heckman--Opdam $\gl_n$ hypergeometric function, see Section 6. 

	Now define Baxter operators $Q_n(\l)$ and raising operators $\Lambda_n(\l)$ similarly to \rf{r9}, \rf{r14}:
	\begin{align}\label{s6}
		& \left( Q_{n}(\lambda) f\right) (\bm{x}_n) =  d_n(g)\, \int_{\mathbb{R}^n} d\bm{y}_n \, Q(\bm{x}_n, \bm{y}_n; \lambda) f(\bm{y}_n), \\[3pt]
		& \label{s7} \left(\cL_{n}(\l)f\right)(\bx_n)= d_{n - 1}(g) \int_{\R^{n-1}}d\by_{n-1} \, \cL(\bx_n,\by_{n-1};\l) f(\by_{n-1}),
	\end{align}
	where the kernels are given by
	\begin{align}\label{s8} 
		& Q(\bx_n,\by_{n};\l) = e^{\const \l(\bbx_n-\bby_n)}
	K(\bx_n,\by_{n})\mu  (\by_{n}),\\[6pt]
		& \cL(\bx_n,\by_{n-1};\l) = e^{\const \l(\bbx_n-\bby_{n-1})}
	\cK(\bx_n,\by_{n-1})\cmu  (\by_{n-1}), \label{s9}
	\end{align}	
	and normalizing constant reads
	\begin{align}
		d_n(g) = \frac{\left(2\pi \Gamma(g)\right)^n}{n!}.
	\end{align}
 Baxter operators commute with Hamiltonians $\H_k$, given by the relations \rf{s3}--\rf{s5}:
\beq\label{a9a} \H_kQ_n(\l)=Q_n(\l)\H_k.
\eeq
The proof of \rf{a9a} copies the proof of analogous identity for the hyperbolic Ruijsenaars system, see e.g. \cite[Section 2]{BDKK1}.
It is a direct consequence of kernel function identity \cite[eq. (35)]{HR0}:
\beq\label{a9b} \H_k(\bx_n) K(\bx_n,\by_n) = \H_k(-\by_n) K(\bx_n,\by_n),\eeq
and of the symmetricity of Hamiltonians $\H_k$ with respect to the bilinear pairing 
\beq \label{a9c}\left(f(\bx_n),\, g(\bx_n)\right)_{\mu}=\int_{\R^n} f(\bx_n)g(-\bx_n)\mu(\bx_n) d\bx_n, \eeq
 which follows from the fact that operators~$\mathcal{H}_k$ given by~\eqref{s4} are manifestly symmetric with respect to the pairing with measure~$d\bx_n$. Note also that commutativity relation~\eqref{a9a} on the space $L^2(\mathbb{R}^n, \mu)$ is established in~\cite[Section~4]{H}.

Using a degeneration of kernel function identity \rf{a9b}, M. Halln\"as and S. Ruijsenaars proved in \cite{HR0} that the function
\beq \label{s10}\Psi_{\bl_n}(\bx_n)=\cL_{n}(\l_n) \, \cL_{n-1}(\l_{n-1})\cdots \cL_{2}(\l_2) \, e^{\const \l_1x_1}
\eeq
is a joint eigenfunction of Hamiltonians \rf{s3}
 \beq\label{s10-2} \H_k \Psi_{\bl_n}(\bx_n)=  e_k( 2\pi \l_1,\ldots, 2\pi \l_n) \Psi_{\bl_n}(\bx_n).\eeq
 Recently, M. Halln\"as \cite{H} proved that the functions \rf{s10} also diagonalize Baxter $Q$-operators \rf{s6}:
\beq\label {s11}
	Q_n(\lambda) \, \Psi_{\bm{\lambda}_n}(\bx_n) = \prod^n_{i=1}\ccK(\lambda-\lambda_i) \, \Psi_{\bm{\lambda}_n}(\bx_n),
\eeq
where 
\begin{equation}\label{s12}
	\ccK(\l) = \Gamma \Bigl(\imath\l + \frac{g}{2} \Bigr) \Gamma\Bigl(-\imath\l+\frac{g}{2}\Bigr).
\end{equation}
 The proof given in \cite{H} relies on the well developed theory of Heckman--Opdam hypergeometric functions.  Below we present an alternative approach, studying the limit of the corresponding construction in relativistic case~\eqref{r21}.

\subsection{Calogero--Sutherland limit} \label{section3.1}
In this section we establish local relations between Baxter and raising operators by using the nonrelativistic limit of Ruijsenaars hyperbolic system. As a corollary, we obtain alternative proof of the diagonalization property~\eqref{s11}. The present approach gives new nontrivial integral identities corresponding to local relations between Baxter and raising operators. 

Consider the Ruijsenaars hyperbolic system with fixed period $\o_1$ and renormalized coupling constant $\rg$
\beq\label{s13}
 \o_1 \mapsto 1,\qquad \rg \mapsto\cg\o_2. \eeq
Let us describe the limits of kernel and measure functions, Baxter operators \rf{r9}, raising operators \rf{r14} and wave functions \rf{r16} (without normalization \rf{dconst}) as $\o_2 \to 0$.
The limiting procedure is based on the following statements, proved in Appendix~\ref{AppB}.

\begin{proposition}\label{proposition3.1} For $g>0$, $x\in \R$ we have pointwise limits 
	\begin{align}\label{s14} 
		& \lim_{\o_2\to 0} \rK(x|1,\o_2,g\o_2)=K(x),\\ \label{s15}
		& \lim_{\o_2\to 0} \rmu(x|1,\o_2,g\o_2)=\mu(x).	
	\end{align}
	\end{proposition}
\begin{proposition}\label{proposition3.2} For $g > 0$ there exist constants $C_1(g,a), C_2(g,a)$ such that
	\begin{align}\label{s16} 
		& |\rK(x|1,\o_2,g\o_2)|<C_1e^{-\pi g|x|},\\[3pt]
		\label{s17} 
		& |\rmu(x|1,\o_2,g\o_2)|< C_2e^{2\pi g|x|}\end{align}
	uniformly for all $x\in\R$ and $\o_2\in (0,a]$. 
	\end{proposition}
The statements of Proposition \ref{proposition3.1} are well known; they easily follow from integral representations of logarithms of double sine and sine functions. However the statements of Proposition \ref{proposition3.2}, especially the bound \rf{s17}, are not so obvious. They are crucial for interchanging limits and integrals, representing kernels of operators and of the wave functions.
The proof is given in Appendix \ref{AppB}. 

For any $2n$ tuple 
$$\bz_{2n}= \{z_1,\ldots, z_n,x_1,\ldots, x_n\}$$
of real numbers, set
\begin{align}\label{J4}  
	& Q_n^R ( \bz_{2n}; \l) = \int_{ \R^n}e^{\const\l\bby_n}\prod_{a=1}^{2n}\prod_{i=1}^n K^R(z_a-y_i) \mu^R(\by_n)d\by_n, \\[6pt] \label{J42}  
	& Q_{n}( \bz_{2n}; \l) = \int_{ \R^n}e^{\const\l\bby_n}\prod_{a=1}^{2n}\prod_{i=1}^n K(z_a-y_i) \mu(\by_n)d\by_n .
\end{align}
 In \cite[Section 3]{BDKK1} the integral identity
\begin{align} \label{J4-2}
	Q^R_{n}(\bz_{2n}; \l)=e^{\const\l\bbz_{2n}}Q^R_{n}(\bz_{2n}; -\l )
\end{align}
is proven under assumption $|\Im \lambda| < \Re \rg/\o_1 \o_2 = g$. This identity is equivalent to commutativity of relativistic Baxter operators~\eqref{r12} written in terms of kernels.
The following statement is its nonrelativistic limit. 
\begin{proposition}\label{proposition3.3} Assuming $| \Im \l | < g$ we have the equality
\beq\label{J5}
Q_{n}(\bz_{2n}; \l)=e^{\const\l\bbz_{2n}}Q_{n}(\bz_{2n}; -\l ).
\eeq
\end{proposition}
{\bf Proof}.  The stated equality follows from the formula~\eqref{J4-2} and the fact that
\begin{align}\label{limQRQ}
	\lim_{\omega_2 \to 0} Q_n^R(\bz_{2n}; \l) = Q_n(\bz_{2n};\l),
\end{align}
which we prove now. By Proposition~\ref{proposition3.1} the integrand of~\eqref{J4} tends to the integrand of~\eqref{J42} as $\omega_2 \to 0$. Moreover, by Proposition~\ref{proposition3.2} the integrand of~\eqref{J4} is bounded by integrable function independent of $\omega_2$ 
\begin{align*}
	\biggl| e^{\const\l\bby_n}&\prod_{a=1}^{2n}\prod_{i=1}^n K^R(z_a-y_i) \mu^R(\by_n) \biggr| \leq C(\bz_{2n}; g) \, \exp \biggl( 2\pi (| \Im \l| - g ) \sum_{j = 1}^n | y_j|   \biggr).
\end{align*}
Hence, due to dominated convergence theorem we can interchange the limit $\omega_2 \to 0$ with integration and arrive at the formula~\eqref{limQRQ}. 
 \hfill{$\Box$} \medskip

 The last proposition gives an alternative to \cite{H} proof of commutativity of nonrelativistic Baxter operators.
 
 \begin{theorem}\label{theorem3.1} Baxter operators commute
 	\beq\label{s18} Q_n(\l)Q_n(\rho)=Q_n(\rho)Q_n(\l)\eeq
 	under assumption $| \Im (\l - \rho) | < g$. 
 	\end{theorem} 
 Indeed, the relation \rf{J5} establishes the equality of kernels of integral operators, representing products of operators
  in both sides of \rf{s18}.
 At the same time, writing the following equality in terms of kernels
\begin{align}\label{s19} 
	\Lambda_n(\l)\Lambda_{n-1}(\rho)=\Lambda_n(\rho)\Lambda_{n-1}(\l), 
\end{align}
one can see that it is also equivalent to the relation \rf{J5}. The above exchange relation between raising operators implies the symmetry 
  	\begin{align*} 
  		\Psi_{\bl_n}(\bx_n)=\Psi_{\sigma(\bl_n)}(\bx_n),\qquad \sigma\in S_n. \end{align*}	
   
As well as for the Ruijsenaars hyperbolic system, the relation \rf{s18} admits a degeneration, which leads to an exchange relation between Baxter and raising operators (recall the function $\ccK(\l)$ given by the formula~\rf{s12}).

 \begin{proposition}\label{proposition3.5} Assuming $|\Im(\lambda-\rho)| < \frac{g}{2}$ we have
	\beq\label{s20} Q_n(\l)\Lambda_n(\rho)=\ccK(\l-\rho) \, \Lambda_n(\rho)Q_{n-1}(\l). \eeq
	\end{proposition} 
 
{\bf Proof}. In \cite{BDKK2}, the relation \rf{s20}, treated as an identity of kernels of corresponding integral operators, is derived by means of asymptotical analysis  of the
integral identity \rf{J5} when one of the parameters, say $z_{2n}$, tends to infinity. The derivation  uses: 
\begin{itemize}  \item[(a)] asymptotics for $ x\rightarrow \pm \infty$
\begin{equation}\label{IV8}
	\rmu(x) \sim e^{2\pi \frac{\rg}{\o_1\o_2} | x | }, \qquad  \rK(x) \sim e^{- \pi\frac{\rg}{\o_1\o_2}  |x|};
\end{equation}
\item[(b)] bounds for $ x \in \mathbb{R}$
\begin{equation}\label{IV16-0}
	| \rmu  (x) | < C_1 e^{2\pi\nu_g |x| }, \qquad |\rK (x)| < C_2 e^{- \pi\nu_g |x| }, 
\end{equation}
where 
\beq \nu_g=\Re \frac{\rg}{\o_1\o_2} >0 ;\eeq 
\item[(c)] Fourier transform formula
\begin{equation*}
	\int_{\mathbb{R}} dy \; e^{2 \pi \imath \lambda y}  \rK (y) = \sqrt{\omega_1 \omega_2} \, S(\rg) \, \rrK (\lambda  ).
\end{equation*}
\end{itemize}
In nonrelativistic Calogero--Sutherland model we have instead:
\begin{itemize}  \item[(d)] asymptotics 
	\begin{equation}
		\mu(x) \sim e^{2\pi {g} | x | }, \qquad K(x) \sim e^{- \pi{g}  |x|}, \qquad x\rightarrow \pm \infty;
	\end{equation}
	\item[(e)] bounds
	\begin{equation}\label{IV16}
		| \mu  (x) | < C'_1 e^{2\pi g |x| }, \qquad |K (x)| < C'_2 e^{- \pi g |x| }, \qquad x \in \mathbb{R};
	\end{equation}
	\item[(f)] Fourier transform formula
	\begin{equation*}
		\int_{\mathbb{R}} dy \; e^{{2 \pi \imath}{}  \lambda y }  K (y) = \frac{\Gamma\left(\imath\l+\frac{g}{2}\right)\Gamma\left(-\imath\l+\frac{g}{2}\right)}{2\pi\Gamma(g)} =\frac{1}{2\pi\Gamma(g)} \ccK (\lambda  ).
	\end{equation*}
\end{itemize}
Replacing in the proof of \cite[Theorem 3]{BDKK2} items (a)--(c) by items (d)--(f), we get the proof of Proposition \ref{proposition3.5}.
\hfill{$\Box$}

As a corollary, we obtain a proof of spectral properties of Baxter operators~\rf{s11}.  The subtle point is the convergence of all integrals appearing on the way, however, the arguments for this are identical to those in relativistic case, see~\cite[Section 3.2]{BDKK2}. 

\section{Ruijsenaars rational system} \label{sec:rat-ruij}
	\subsection{Statements}
		 Denote by $T_{\l_k}$ the shift operators
		 \begin{align*}
		 	T_{\l_k}f(\l_1,\ldots, \l_k,\ldots, \l_n)= f(\l_1,\ldots, \l_k -\i ,\ldots, \l_n).
		 \end{align*}
		Commuting Hamiltonians for rational Ruijsenaars system in Macdonald form look as
		\beq\label{g1}
		\ccM_r:=\ccM_r(\bl_n)=
		(-1)^{r(n-1)}\sum_{\substack{I\subset[n] \\ |I|=r}}
		\prod_{\substack{i\in I \\ j\notin I}}
		\frac{\l_{i}-\l_{j}  - \i (1 - g)}{\l_{i}-\l_{j}}
		\prod_{i\in I}T_{\l_{i}}.
		\eeq
	 In this section and in what follows we assume
	 \begin{equation}
	 	g > 1.
	 \end{equation}
	  All statements below are in full analogy with the hyperbolic Ruijsenaars system with double sine functions replaced by gamma functions. To prove these statements we follow the same steps, as for the hyperbolic model~\cite{BDKK1, BDKK2}.
	  
	  \begin{remark}
	  	Recall that the hyperbolic wave functions diagonalize the operators~\eqref{MR-dual}
	  	\begin{align} 
	  		\hat{M}^R_r = \sum_{\substack{I\subset[n] \\ |I|=r}} \prod_{\substack{i\in I \\ j\notin I}} \frac{\sh \pi \o_2 \bigl(\l_i-\l_j-\imath \frac{\omega_1 + \omega_2 - \rg}{\omega_1 \omega_2} \bigr)}{\sh \pi \o_2(\l_i-\l_j)} \cdot T^{-\imath/\o_1}_{I,\l}.
	  	\end{align}
	  	The latter tend to the rational Hamiltonians~\eqref{g1} in the nonrelativistic limit from Section~\ref{section3.1}: taking $\omega_1 = 1$, $\rg = g \omega_2$ and $\omega_2 \to 0$.
	  	This is why we consider the rational Ruijsenaars system with reflected coupling constant $1 - g$. Besides, let us note that the rational Hamiltonians with constants $g$ and $1-g$ are related by similarity transformation, see~\eqref{MH-sim}.
	  \end{remark}
	 
	 Now we introduce rational analogs of weight, measure and kernel functions, see the limiting formulas~\eqref{Kw-dual-lim}. Set
		\begin{align}\label{g2} 
			\ccm(\l)= \Gamma^{-1}(\imath \l)\Gamma^{-1}(-\i \l+g),
			\qquad \ccmu(\l)=\ccm(\l)\ccm(-\l)
		\end{align}
			and correspondingly
			\beq\label{g4-0} \begin{split}
				& \ccm(\bl_n)=\prod_{1 \leq i<j \leq n}\ccm(\l_i-\l_j),\qquad	\ccmu(\bl_n)=\prod_{1 \leq i < j \leq n}\ccmu(\l_i-\l_j),\\[3pt]
			& \ccK(\bl_n,\bg_m)=\prod_{i = 1}^n \prod_{a = 1}^m\ccK(\l_i-\gamma_a),\end{split}\eeq
			where the function $\ccK(\l)$ is defined by the formula \rf{s12}. Set also 
			\beq\label{g4}
			 \dd_n(g)=\frac{1}{n!}\left({2\pi} \Gamma(g)\right)^{-n}.\eeq
			In the same way, as for hyperbolic model, we introduce a family of Baxter operators $\ccQ_{n}(x)$ parameterized by $x \in \mathbb{C}$.  These are integral operators
			\begin{equation}\label{g5}
				\bigl( \ccQ_{n}(x) f \bigr) (\bl_n) =  \dd_n(g) \, \int_{\mathbb{R}^n} d\bg_n \, \ccQ(\bl_n, \bg_n; x) f(\bg_n)
			\end{equation}
			with the kernel
			\beq\label{g6} \ccQ(\bl_n,\bg_{n};x)= e^{\const x(\bbl_n-\bbg_n)}
			\ccK(\bl_n,\bg_{n})\ccmu  (\bg_{n}).
			\eeq
			All of the following theorems are proven in Section~\ref{sec:rat-ruij-proofs}.
			\begin{theorem}\label{theoremg1}
				The rational Baxter operators commute with the rational Ruijsenaars--Macdonald operators
				\beq\label{g7-1} \ccM_r\ccQ_{n}(x)=\ccQ_{n}(x)\ccM_r. \eeq
				\end{theorem}
				\begin{theorem}\label{theoremg2}
				The rational Baxter operators commute 
				\beq\label{g7} \ccQ_{n}(x)\ccQ_{n}(y)=\ccQ_{n}(y)\ccQ_{n}(x) \eeq
				 under assumption $|\Im (x - y)| < 1$.
			\end{theorem}
		Similarly, we define raising operators  $\ccL_{n}(x)$ parameterized by $x \in \mathbb{C}$.  These are integral operators, which transform functions of $n-1$ variables to functions of $n$ variables
		\begin{equation}\label{g8}
			\bigl( \ccL_{n}(x) f\bigr) (\bl_n) =  \dd_{n-1}(g)  \int_{\mathbb{R}^{n-1}} d\bg_{n-1} \, \ccL(\bl_n, \bg_{n-1}; x) f(\bg_{n-1})
		\end{equation}
		with the kernel
		\beq\label{g9} \ccL(\bl_n,\bg_{n-1};x)= e^{\const x(\bbl_n-\bbg_{n-1})}
		\ccK(\bl_n,\bg_{n-1})\mm  (\bg_{n-1}) .
		\eeq
		Also recall the definition of kernel function in Calogero--Sutherland model
		\beqq \cK(x)=( 2 \ch \pi x)^{-\cg} .\eeqq
		In the limit from commutativity~\eqref{g7} we obtain the following exchange relation.
			\begin{theorem}\label{theoremg3}  Assuming $| \Im(x - y)| < \tfrac{1}{2}$ we have
		\beq\label{g10}		\ccQ_n(x)\ccL_{n-1}(y)=\cK(x-y) \, \ccL_{n-1}(y)\ccQ_{n-1}(x). \eeq 
	\end{theorem}
	Define the wave function $\ccPsi_{\bl_n}(\bx_n)$ by the formula
	\beq \label{g11} {\ccPsi}_{\bl_n}(\bx_n)=\ccL_n(x_n) \cdots \ccL_2(x_2) \, e^{2\pi \imath \l_1x_1}.\eeq
	These functions diagonalize both rational Ruijsenaars--Macdonald and Baxter operators. 
	\begin{theorem}\label{theoremg4} For  $|\Im x | < \tfrac{1}{2}$  and $\bx_n\in\mathbb{R}^n$ we have
		\begin{align}\label{g12} 
			& \ccM_r \, {\ccPsi}_{\bl_n}(\bx_n)=  e_r(e^{2\pi x_1},\ldots, e^{2\pi x_n}) \, {\ccPsi}_{\bl_n}(\bx_n),\\[6pt]
			\label{g13} 
			& \ccQ_n(x) \, {\ccPsi}_{\bl_n}(\bx_n)=\prod_{j=1}^n \cK(x-x_j) \, {\ccPsi}_{\bl_n}(\bx_n).
			\end{align}
		\end{theorem} 
	Here $e_r(z_1,\ldots,z_n)$ is $r$-th elementary symmetric function. 
	 Note that in~\cite{KK1} it is shown that functions $\ccPsi_{\bl_n}(\bx_n)$ also diagonalize Calogero--Sutherland Hamiltonian.  
	
	As in previous sections, the integral identity related to commutativity of Baxter operators~\eqref{g7} is also related to the exchange relation between raising operators.
	\begin{theorem} \label{theorem5}
		The raising operators satisfy 
		\begin{align}
			\ccL_n(x) \ccL_{n - 1}(y) = \ccL_n(y) \ccL_{n - 1}(x) 
		\end{align}
		under assumption $|\Im (x - y)| < 1$.
	\end{theorem}
	This leads to the symmetry of wave functions
	\begin{align} \label{rat-wf-sym}
		\ccPsi_{\bl_n}(\bx_n) = \ccPsi_{\bl_n}(\sigma(\bx_n)), \qquad \sigma \in S_n.
	\end{align}

\subsection{Proofs} \label{sec:rat-ruij-proofs}	
{\bf 1}. The proof of Theorem \ref{theoremg1} repeats the  corresponding  arguments for hyperbolic  Ruijsenaars system \cite{BDKK1}. It follows  from the rational version of the kernel function  identity~\cite{Ru3}
\begin{equation}\label{g15}\begin{split}
		& \sum_{I\subset[n],~|I|=r}\prod_{i\in I}\left(\prod_{j\notin I}\frac{z_i-z_j-\alpha}{z_i-z_j}\prod_{a = 1}^n \frac{z_i-y_a+\alpha}{z_i-y_a}\right) \\
		& = \sum_{A\subset[n],~|A|=r}\prod_{a\in A}\left(\prod_{b\notin A}\frac{y_a-y_b+\alpha}{y_a-y_b}\prod_{i = 1}^n\frac{y_a-z_i-\alpha}{y_a-z_i}\right)
\end{split}\end{equation}
 and symmetry of rational Ruijsenaars--Macdonald operators with respect to the measure~$\ccmu(\bl_n)$. \medskip
 
\noindent {\bf 2}. The proof of Theorem \ref{theoremg2} is a simplified version of the proof of corresponding statement for 
 hyperbolic system given in \cite{BDKK1}. Just as in \cite{BDKK1} it reduces to the proof of integral identity 
 \beq\label{g16}
 \ccQ_{n}(\bz_{2n}; \l)=e^{\const\l\bbz_{2n}}\ccQ_{n}(\bz_{2n}; -\l ),
 \eeq
where
\beq\label{g17} \ccQ_{n}( \bz_{2n}; \l) = \int_{ \R^n}e^{\const\l\bby_n}\prod_{a=1}^{2n}\prod_{i=1}^n \ccK(z_a-y_i) \ccmu(\by_n)d\by_n .\eeq
 It is not difficult to verify the convergence of both integrals by the use of Stirling formula. Moreover, for $\l$ with big enough imaginary part both integrals can be calculated by summing the residues.  To verify that the integrals over big enclosing contours vanish one needs an estimate for the gamma function near its poles, which can be obtained using Euler reflection formula, e.g. see \cite[Section 2.4]{AAR}.
 
  The residue calculations also simplify since zeroes of the measure function annihilate multiple poles.
  The integral identity \rf{g16} is then reduced to the rational hypergeometric identity
  	\beq\label{g18}\begin{split} 
  		& \sum_{|\bk|=K}\prod_{{i,j=1  }}^n\frac{(x_i-x_j-k_j-\a)_{k_i}}{(x_i-x_j-k_j)_{k_i}}\prod_{a,j=1}^n\frac{(x_j-y_a+\a)_{k_j}}
  	{(x_j-y_a)_{k_j}}\\
  	& = \sum_{|\bk|=K}\prod_{{a,b=1 }}^n
  	\frac{(y_a-y_b-k_a-\a)_{k_b}}{(y_a-y_b-k_a)_{k_b}}\prod_{j,a=1}^n\frac{(x_j-y_a+\a)_{k_a}}
  	{(x_j-y_a)_{k_a}},
  \end{split}\eeq
  see \cite{HLNR, BDKK4}.
Here the sum is taken over all $n$ tuples $\bk=(k_1,\ldots,k_n)$ of nonnegative integers such that 
$\sum_jk_j=K$. \medskip

\noindent {\bf 3}. To prove Theorem~\ref{theoremg3} we follow the arguments of \cite{BDKK2}. For that we first establish necessary asymptotics for dual kernel and measure functions by using well known asymptotics of gamma function when $x\to \infty, \ |\arg x|\le \pi-\delta$:
\beqq \log \Gamma(x)=\frac{1}{2}\log 2\pi +\left(x-\frac{1}{2} \right)\log x-x+O\left(\frac{1}{x}\right) .\eeqq
 Namely, we have
 \begin{align}\label{g19} \ccK(\l)&=2\pi |\l|^{g-1}e^{-\pi|\l|}\left(1+O\left(\frac{1}{\l}\right)\right),\qquad \l\to\infty, \ \l\in\R; \\
 	\label{g20}  {\ccm}^{-1}(\l)&=2\pi |\l|^{g-1}e^{-\pi|\l|}\left(1+O\left(\frac{1}{\l}\right)\right),\qquad \l\to\infty, \ \l\in\R .
 \end{align}
Basic commutativity relation \rf{g7} is an integral identity of kernels \rf{g16} which we rewrite now in a form 
\begin{equation}\label{g20-2}
	\int_{\R^n}\ccQ(\bl_n, \bg_n; x)\ccQ(\bg_n, \bnu_n;  y)d\bg_n =
	\int_{\R^n}\ccQ(\bl_n, \bg_n; y)\ccQ(\bg_n, \bnu_n;  x)d\bg_n.
\end{equation}
Multiply \rf{g20-2} by the function
\beq\label{g21} r(\bnu_n,y)=\nu_n^{(n-2)(g-1)}\exp\Bigl( \pi  \bigl[\bbnu_{n - 1}+(2-n+2\imath y)\nu_n\bigr] \Bigr),\eeq
\begin{equation}\label{g20a}
	r(\bnu_n,y)\int\limits_{\R^n}\ccQ(\bl_n, \bg_n; x)\ccQ(\bg_n, \bnu_n;  y)d\bg_n =
r(\bnu_n,y)	\int\limits_{\R^n}\ccQ(\bl_n, \bg_n; y)\ccQ(\bg_n, \bnu_n;  x)d\bg_n\end{equation}
and take the limit $\nu_n \to +\infty$.
 Then, due to \rf{g19}, \rf{g20}, the integrand in the left hand side of \rf{g20a} has pointwise limit equal to 
\beq\label{g22} (2\pi)^{2-n}\ccQ(\bl_n, \bg_n; x)\ccL\left(\bg_n, \bnu_{n-1};  y-\frac{\imath}{2}\right) .\eeq
 To pull the limit inside the integral we use dominated convergence theorem. This requires estimating the integrand by integrable function independent of $\nu_n$.
 To estimate the integrand, we use the bounds
 \begin{align}\label{g19a}
 & |\ccK(\l)|<C\left(\l^2+1\right)^{\frac{g-1}{2}}e^{-\pi|\l|},\\
 	 & |\ccm(\l)|<C\left(\l^2+1\right)^{\frac{1-g}{2}}e^{\pi|\l|},\label{g19b}\end{align}
  which follow from \rf{g19}, \rf{g20}. Note also that under the condition $ g>1 $  inequalities \rf{g19a} and \rf{g19b} imply 
   \begin{align}\label{g19c}
   	& |\ccK(\l)|<p(|\l|)e^{-\pi|\l|},\\
  \label{g19d} & |\ccm(\l)|<C'e^{\pi|\l|}, \end{align}
  where $p(|\l|)$ is some polynomial with positive coefficients, whose degree depends on $g$. They will be also exploited further.
  
  The calculation of exponential part of the bound repeats that of \cite{BDKK2} and gives the exponent
  \beq\label{g23} C e^{ 2\pi\left(\Im\left(x-y+\frac{\imath}{2}\right)-1\right)\sum_j|\g_j|},\eeq
  where the constant depends on fixed parameters $\bl_n$, $\bnu_{n-1}$.
The power part of the bound is equal to the product
\beq\label{g24} F_1(\bg_n)\cdot F_2(\bg_n,\nu_n)\cdot F_3(\nu_n),\eeq
where
\beqq\begin{split} F_1(\bg_n)=\prod_{i,j=1}^n\left((\l_i-\g_j)^2+1\right)^{\frac{g-1}{2}} 
\prod_{1\leq i<j\leq n}\left((\g_i-\g_j)^2+1\right)^{1-g}\; \prod_{j=1}^n\prod_{k=1}^{n-1}\left((\nu_k-\g_j)^2+1\right)^{\frac{g-1}{2}},\end{split}
\eeqq
\beqq\begin{split}
 F_2(\bg_n,\nu_n)=&\prod\limits_{j=1}^n\left(\frac{(\g_j-\nu_n)^2+1}{\nu_n^2}\right)^{\frac{g-1}{2}},\qquad
 F_3(\nu_n)= \prod\limits_{j=1}^{n-1}\left(\frac{(\nu_j-\nu_n)^2+1}{\nu_n^2}\right)^{1-g}.
\end{split}	\eeqq
The function $F_1(\bg_n)$ is a smooth positive function of power growth and thus can be bounded by an exponect
\beqq C\exp(\delta\sum_j|\g_j|) \eeqq
with arbitrary positive $\delta$ and a proper constant $C$. 

For $g>1$, the  function $ F_2(\bg_n,\nu_n)$ is a product of positive powers of the factors
$$ \left(1-\frac{\g_j}{\nu_n}\right)^2+\frac{1}{\nu_n^2},$$
and consequently, can be bounded for arbitrary $\delta>0$ by an exponent
\beqq C(\nu_n)\exp(\delta\sum_j|\g_j|), \eeqq
 where the constant $C(\nu_n)$ is restricted from above when $\nu_n$ tends to infinity.
 Finally, the function $F_3(\nu_n)$ with fixed $\nu_j$ ($j = 1, \dots, n - 1$) tends to $1$ when $\nu_n$ tends to infinity and as a result can be restricted by a constant for big $\nu_n$.
 In total, the integrand from the left hand side is bounded by
 \beq C \exp\left({ 2\pi\left(\Im\left(x-y+\frac{\imath}{2}\right)-1+\delta\right)\sum_j|\gamma_j|}\right)\eeq
 with arbitrary small positive $\delta$. Thus, if
 \beq \label{g25}-1< \Im (x-y)<0 \eeq
 the integrand is bounded by integrable function uniformly in $\nu_n$. Then we can apply dominant convergence theorem and state that the left hand side of \rf{g20a} has a limit equal to
 \beq\label{g26} (2\pi)^{2-n}\int\limits_{\R^n}\ccQ(\bl_n, \bg_n; x)\ccL\left(\bg_n, \bnu_{n-1};  y-\frac{\imath}{2}\right)d\bg_n .\eeq

In the right hand side of \rf{g20a}, using the symmetry of the integrand, we replace the integration over $\R^n$ by $n$ times integral
 over the region $\g_j<\g_n$ for all $j$ and then perform the change of variable $\g_n\to\g_n+\nu_n$. After that, we can see that the integrand also has a pointwise limit equal to
 \beq\label{g26a}\begin{split} & n(2\pi)^{2-n}e^{2\pi \imath \left[\left(y-\frac{\imath}{2}\right)(\bbl_n-\bbg_n)+x(\bbg_n-\bbnu_{n-1})\right]}\\
 	& \times\ccK(\bl_n,\bg_{n-1})\ccmu(\bg_{n-1})\ccK(\bg_{n-1},\bnu_{n-1})\ccmu(\bnu_{n-1})\ccK(\g_n). 
 	\end{split}\eeq
  Again, to perform the limit we use dominated convergence theorem. The exponential bound of the integrand coincides with that given in \cite{BDKK2} after  proper change of parameters and can be chosen as in \rf{g23}. The power part can be presented as a product
 \beqq  G_1(\bg_n)\cdot G_2(\bg_n,\nu_n)\cdot G_3(\nu_n)\cdot G_4(\bg_n,\nu_n),\eeqq
 where
 \beqq\begin{split} G_1(\bg_n)=\prod_{i=1}^n\prod_{j=1}^{n-1}\left((\l_i-\g_j)^2+1\right)^{\frac{g-1}{2}} 
 \prod_{j=1}^{n-1}\prod_{k=1}^{n-1}\left((\nu_k-\g_j)^2+1\right)^{\frac{g-1}{2}}\\ 
\times \prod_{1\leq i<j\leq n-1}\left((\g_i-\g_j)^2+1\right)^{1-g}\,\left(\g_n^2+1\right)^{1-g}\!\!,\end{split}\eeqq
 \beqq\begin{split}
 	G_2(\bg_n,\nu_n)=&\prod\limits_{j=1}^{n-1}\left(\frac{(\g_j-\nu_n)^2+1}{\nu_n^2}\right)^{\frac{g-1}{2}}\,
 \left(\frac{(\l_n-\g_n-\nu_n)^2+1}{\nu_n^2}\right)^{\frac{g-1}{2}}	,\\
 	G_3(\nu_n)= &\prod\limits_{j=1}^{n-1}\left(\frac{(\nu_j-\nu_n)^2+1}{\nu_n^2}\right)^{1-g},\\
 	G_4(\bg_n,\nu_n)=&\prod_{j=1}^{n-1}\left(\frac{\left((\l_j-\g_n-\nu_n)^2+1\right)\left((\nu_j-\g_n-\nu_n)^2+1\right)}{\left((\g_j-\g_n-\nu_n)^2+1\right)^2}\right)^{\frac{g-1}{2}}.
 \end{split}	\eeqq
 The products $G_1(\bg_n)$, $G_2(\bg_n,\nu_n)$ and  $G_3(\nu_n)$ are bounded by the same arguments as $F_1(\bg_n)$, $F_2(\bg_n,\nu_n)$ and  $F_3(\nu_n)$ above.
  For the last product $	G_4(\bg_n,\nu_n)$, we use the  inequality
  \beq\label{g26b}
  \frac{(x-a)^2+1}{(x-b)^2+1}<(a-b)^2+2 \eeq
  which follows from the analysis of the elementary function of real variable
  \beqq f(x)=\frac{(x-a)^2}{x^2+1}. \eeqq
  Inequality \rf{g26b} implies that the product 	$G_4(\bg_n,\nu_n)$ can be bounded by positive power function of the variables $\bg_n$, which does not depend on $\nu_n$ and thus can be suppressed by an exponent 
  \beqq C\exp(\delta\sum_j|\g_j|)\eeqq
   with any positive $\delta$.
 
 We thus conclude that under condition \rf{g25} we have a uniform bound of the integrand by the integrable function and can interchange limit and integral.
  The integral over $\g_n$ separates, 
  \beq\label{g27}\int_{\R}d\g_n \, e^{2\pi \imath\left(x-y+\frac{\imath}{2}\right)\g_n}\Gamma\left(\imath\gamma_n+\frac{g}{2}\right)
  \Gamma\left(-\imath\gamma_n+\frac{g}{2}\right)= 2\pi\Gamma(g)\left(2\ch\pi\left(x-y+\frac{\imath}{2}\right)\right)^{-g},\eeq
  that is
  \beq\label{g28} \int_{\R}d\g_n \, e^{2\pi \imath\left(x-y+\frac{\imath}{2}\right)\g_n}\ccK(\g_n)=
  2\pi\Gamma(g)\cK\left(x-y+\frac{\imath}{2}\right).\eeq
  Combining \rf{g28} with \rf{g26}, \rf{g22}, and \rf{g4}, we come to the relation
  \begin{equation} \label{g29}
  	\ccQ_n(x) \, \ccL_{n}\Bigl(y - \frac{\imath }{2} \Bigr) = \cK \Bigl(x - y + \frac{\imath }{2} \Bigr) \ccL_{n}\Bigl(y - \frac{\imath }{2} \Bigr) \ccQ_{n-1}(x),
  \end{equation}
valid under condition \rf{g25}. Thus we can make a shift $y\to y+\frac{\imath}{2}$ and get the statement \rf{g10} of Theorem \ref{theoremg3}. \medskip

\noindent {\bf 4.} 	The first part \rf{g12} of Theorem~\ref{theoremg4} was proved in \cite{KK}. The second part is a direct corollary of Theorem \ref{theoremg3}. \medskip

\noindent {\bf 5.} In terms of kernels the relation stated in Theorem~\ref{theorem5} is equivalent to the already proven identity~\eqref{g16}.

\setcounter{equation}{0}
\section{Asymptotics and duality} \label{sec:as-dual}
\subsection{Asymptotics and bounds} \label{sec:as}
	In  \cite{HR2} and \cite{HR3} M. Halln\"as and S. Ruijsenaars got remarkable results on asymptotical behaviour and total bounds of the wave functions of hyperbolic Ruijsenaars model. These bounds were used in \cite{BDKK2} for establishing the duality symmetry~\eqref{r26} of the wave functions of hyperbolic Ruijsenaars model.
	
	The wave function ${\ccPsi}_{\bl_n}(\bx_n)$ given by iterative Mellin--Barnes integral \rf{g11} admits analogous  asymptotical analysis with arguments almost the same, as in \cite{HR2, HR3}.
	
	Following the lines of \cite{HR2, HR3}, let us introduce renormalized wave functions 
	\beq\label{g32} \Phi_{\bl_n}(\bx_n)=\ccm(\bl_n){\ccPsi}_{\bl_n}(\bx_n) \eeq
	and (as shown below) their asymptotics
	\beq\label{g33}\Phi^{as}_{\bl_n}(\bx_n)= e^{-2 \pi  g(\bx_n,\brho_n)}\sum_{\sigma\in S_n} e^{2\pi \imath \sum_{i = 1}^n x_i\l_{\sigma(i)}}\!\!\! \prod_{\substack{i<j\\ \sigma(i)>\sigma(j)}}\frac{\ccm(\l_{\sigma(j)}-\l_{\sigma(i)})}{\ccm(\l_{\sigma(i)}-\l_{\sigma(j)})},\eeq
	where 
	\beq\label{grho} \brho_n=\Big(\frac{n-1}{2},\ldots, \frac{1-n}{2}\Big).\eeq
	Set also \beq\label{g33c}
	\ddelta(\bl_n)=\prod_{1 \leq i<j \leq n} (1+|\l_i-\l_j|).\eeq
	Then we have the following statements. Assume that 
	\beq\label{g33b} x_1>x_2>\ldots >x_n \eeq
	and denote by $ m_n(\bx_n)$ the minimum of distances $x_i-x_{i+1}$,
	\beq\label{gmn} m_n(\bx_n) =\min_i (
x_i-x_{i+1}). \eeq 
\begin{theorem}\label{theorem5.1}
	For $g>1$, $\bl_n \in \mathbb{R}^n$  and $\ve \in (0, 2\pi)$ there exist constants $N(g), C(g, \ve)>0$ such that in the region \rf{g33b}
	\beq\label{g34a} |\Phi_{\bl_n}(\bx_n)-\Phi^{as}_{\bl_n}(\bx_n)|<C(g,\ve) \, 
	\ddelta(\bl_n)^{N(g)} \, e^{-2 \pi  g(\bx_n,\brho_n)-(2\pi-\ve) m_n(\bx_n) }.\eeq
\end{theorem}
 Since wave function is symmetric with respect to $\bx_n$~\eqref{rat-wf-sym}, the above theorem in fact provides its asymptotics in all regions. Besides, we have a total bound on the function $\Phi_{\bl_n}(\bx_n)$. 
\begin{theorem}\label{theorem5.2}
	For $g>1$  and $\bx_n, \bl_n \in \mathbb{R}^n$  there exist $N(g),C(g)>0$ such that
	\beq\label{g34b} |\Phi_{\bl_n}(\bx_n)|<C(g) \, \ddelta(\bl_n)^{N(g)} \, e^{-\pi g\sum_{i<j}|x_i-x_j|}.
	\eeq 
\end{theorem}
\noindent Note that, actually, in both theorems we can take
\beq N(g) = 4^{n-2}g - \frac{(4^{n-2}-1)(g-2)}{3}. \eeq
 The proofs of these theorems, which are given in Appendix \ref{AppD},  essentially follow the arguments by Halln\"as and Ruijsenaars with the basic bounds on the hyperbolic kernel and weight functions replaced by inequalities \rf{g19c} and \rf{g19d}.

\subsection{Duality} \label{sec:dual}
		\begin{theorem}\label{theoremg5}
        For $g > 1$ and  $\bx_n,\bm{\lambda}_n\in\mathbb{R}^n$ we have the equality
		\beq\label{g14}{\cPsi}_{\bl_n}(\bx_n)={\ccPsi}_{\bl_n}(\bx_n).\eeq
	\end{theorem}
{\bf Proof}. The proof is by induction over $n$. For $n = 1$ the equality is immediate
$$\hat\Psi_{\lambda_1}(x_1) = e^{2\pi\iota \lambda_1x_1} = \Psi_{\lambda_1}(x_1).$$
Now assume we proved the statement for $n - 1$ variables.
By definition (\ref{s10}),
\beq\begin{split}
\Psi_{\bm{\lambda}_n}(\bx_n) = \Lambda_n(\lambda_n)\Psi_{\bm{\lambda}_{n-1}}(\bx_n).
\end{split}\eeq
Comparing (\ref{s6}) and (\ref{s7}), we obtain
\beq\begin{split}
\Psi_{\bm{\lambda}_n}(\bx_n) = e^{2\pi\iota\lambda_n x_n}Q_{n-1}(\lambda_n) \prod^{n-1}_{j=1}K(x_n-x_j)  \Psi_{\bm{\lambda}_{n-1}}(\bx_{n-1}).
\end{split}\eeq
According to (\ref{g13})
\beq\begin{split}
\prod^{n-1}_{j=1}K(x_n-x_j)\hat\Psi_{\bm{\lambda}_{n-1}}(\bx_{n-1}) = \hat Q_{n-1}(x_n)\hat\Psi_{\bm{\lambda}_{n-1}}(\bx_{n-1}),
\end{split}\eeq
so applying induction assumption $\Psi_{\bm{\lambda}_{n-1}}(\bx_{n-1}) = \hat\Psi_{\bm{\lambda}_{n-1}}(\bx_{n-1})$ we deduce
\beq\begin{split}\label{dir_qq}
\Psi_{\bm{\lambda}_n}(\bx_n) = e^{2\pi\iota\lambda_n x_n}Q_{n-1}(\lambda_n)\hat Q_{n-1}(x_n) \Psi_{\bm{\lambda}_{n-1}}(\bx_{n-1}).
\end{split}\eeq
The next step is to change the order of Baxter operators 
$$Q_{n-1}(\lambda_n)\hat Q_{n-1}(x_n) = \hat Q_{n-1}(x_n)Q_{n-1}(\lambda_n).$$
To do it, we have to prove that the integral~\eqref{dir_qq} 
\beq\begin{split}\label{integ}
	\int_{\mathbb{R}^{2(n-1)}} & d\by_{n-1}d\bm{\gamma}_{n-1} \,  e^{2\pi\iota\lambda_n(\underline{\bx}_{n - 1} - \underline{\by}_{n - 1})} \, e^{2\pi\iota x_n( \underline{\bl}_{n - 1} - \underline{\bg}_{n - 1})}\\[6pt] 
	& \times K(\bx_{n-1},\by_{n-1})\hat K(\bm{\lambda}_{n-1},\bm{\gamma}_{n-1})\mu(\by_{n-1})\hat\mu(\bm{\gamma}_{n-1})\Psi_{\bm{\gamma}_{n-1}}(\by_{n-1})
\end{split}\eeq
is absolutely convergent and then apply Fubini's theorem. 

To prove absolute convergence we use the bound on wave function (\ref{g34b}). More precisely, we split the measure functions as
\beqq\begin{split}
\mu(\by_{n-1}) = w(\by_{n-1})w(-\by_{n-1}), \qquad \hat\mu(\by_{n-1}) = \hat w(\by_{n-1})\hat w(-\by_{n-1}),
\end{split}\eeqq
and group factors from the integrand of (\ref{integ}) into three parts
\beq\begin{split}
& w(-\by_{n-1}) K(\bx_{n-1},\by_{n-1}), \qquad \hat w(-\bm{\gamma}_{n-1})\hat K(\bm{\lambda}_{n-1},\bm{\gamma}_{n-1}), \\[6pt]
& w(\by_{n-1})\hat w(\bm{\gamma}_{n-1})\Psi_{\bm{\gamma}_{n-1}}(\by_{n-1})  .
\end{split}\eeq
According to (\ref{IV16}), we have
\beq\begin{split}\label{bndOne}
|w(-\by_{n-1})K(\bx_{n-1},\by_{n-1})| & \leq C^{\prime}_1e^{\pi g\sum_{i < j}|y_i-y_j|}\cdot C^{\prime}_2e^{-\pi g\sum_{i,j}|x_i-y_j|} \\
& \leq C^{\prime}(g,\bm{\omega},\bx_{n-1})e^{-\pi g\sum^{n-1}_{j=1}|y_j|},
\end{split}\eeq
where we use inequalities $|y_i-y_j| \leq |y_i| + |y_j|$ and $|x_i-y_j| \geq |y_j|-|x_i|$.

Similarly, from (\ref{g19c}) and (\ref{g19d}) we deduce
\beq\begin{split}\label{bndTwo}
|\hat w(-\bm{\gamma}_{n-1})\hat K(\bm{\lambda}_{n-1},\bm{\gamma}_{n-1})| \leq \hat C^{\prime}(g,\bm{\omega},\bm{\lambda}_{n-1}) p(|\bm{\gamma}_{n-1}|) e^{-\pi\sum^{n-1}_{j=1}|\gamma_j|}.
\end{split}\eeq
Furthermore, applying (\ref{g34b}) we have
\beq\begin{split}\label{bndThree}
|w(\by_{n-1})\hat w(\bm{\gamma}_{n-1})\Psi_{\bm{\gamma}_{n-1}}(\by_{n-1})| \leq C(g,\bm{\omega})\ddelta(\bm{\gamma}_{n-1})^{N'(g)},
\end{split}\eeq
where we again used induction assumption $\Psi_{\bm{\gamma}_{n-1}}(\by_{n-1}) = \hat\Psi_{\bm{\gamma}_{n-1}}(\by_{n-1})$. Bounds (\ref{bndOne}), (\ref{bndTwo}), (\ref{bndThree}) imply absolute convergence of the integral~\eqref{integ}.

Changing order of $Q$-operators in (\ref{dir_qq}), we have
\beq\begin{split}
\Psi_{\bm{\lambda}_n}(\bx_n) & = e^{2\pi\iota\lambda_n x_n}Q_{n-1}(\lambda_n)\hat Q_{n-1}(x_n) \Psi_{\bm{\lambda}_{n-1}}(\bx_{n-1}) \\[6pt]
& = e^{2\pi\iota\lambda_n x_n}\hat Q_{n-1}(x_n)Q_{n-1}(\lambda_n) \Psi_{\bm{\lambda}_{n-1}}(\bx_{n-1}).
\end{split}\eeq
Using the identity (\ref{s11})
\beq\begin{split}
Q_{n-1}(\lambda_n)\Psi_{\bm{\lambda}_{n-1}}(\bx_{n-1}) = \prod^{n-1}_{j=1}\hat K(\lambda_n-\lambda_j)\Psi_{\bm{\lambda}_{n-1}}(\bx_{n-1}),
\end{split}\eeq
we rewrite the above expression as
\beq\begin{split}
\Psi_{\bm{\lambda}_n}(\bx_n) = e^{2\pi\iota\lambda_n x_n}\hat Q_{n-1}(x_n)K(\lambda_n-\lambda_j)\Psi_{\bm{\lambda}_{n-1}}(\bx_{n-1}).
\end{split}\eeq
Finally, due to explicit formulas (\ref{g6}), (\ref{g9}) and induction assumption we have
\beq\begin{split}
\Psi_{\bm{\lambda}_n}(\bx_n) = \hat \Lambda_n(x_n)\hat\Psi_{\bm{\lambda}_n}(\bx_{n-1}).
\end{split}\eeq
This is exactly the definition of Mellin--Barnes integral representation (\ref{g11}).
\hfill{$\Box$}

\section{Heckman--Opdam hypergeometric function} \label{sec:HO}
 In the previous sections we considered wave functions of Calogero--Sutherland model constructed using integral operators. Alternatively, Heckman and Opdam studied wave functions expressed in terms of so-called Harish-Chandra series~\cite{HO}. In particular, they proved that Calogero--Sutherland Hamiltonians~\eqref{s3} admit unique eigenfunctions $F_{\bl_n}(\bx_n)$ with normalization
\begin{equation}\label{norm}
	F_{\bl_n}(0, \dots, 0) = 1,
\end{equation}
which are symmetric and analytic in $\bx_n$ on a suitable neighbourhood of the origin.

The results of the previous section enable us to establish the precise correspondence between Heckman--Opdam $\mathfrak{gl}_n$ hypergeometric function and the constructed above wave function of the Calogero--Sutherland model.

First, we reformulate the statement of Theorem \ref{theorem5.1} directly for the function  
\beqq {\Psi}_{\bl_n}(\bx_n)  = {\ccPsi}_{\bl_n}(\bx_n) = \frac{1}{\hat{w}(\bl_n)} \, \Phi_{\bl_n}(\bx_n).\eeqq
 Define  
\begin{align} \label{g35} 
	 {\Psi}^{as}_{\bl_n}(\bx_n) =  \frac{1}{\hat{w}(\bl_n)} \, \Phi^{as}_{\bl_n}(\bx_n) = e^{-2 \pi  g(\bx_n,\brho_n)}\sum_{\sigma\in S_n}e^{2\pi \imath \sum_{i = 1}^n x_i\l_{\sigma(i)}} \prod_{{i<j}}\ccm^{-1}(\l_{\sigma(i)}-\l_{\sigma(j)}).
\end{align}
 Theorem \ref{theorem5.1} implies the following  asymptotical behaviour of the function  ${\Psi}_{\bl_n}(\bx_n)$ for real $\bl_n$.
\begin{proposition}\label{proposition5.1}  In the region $x_1 > x_2 > \ldots > x_n$
	\beq\label{g36} |\Psi_{\bl_n}(\bx_n)-\Psi^{as}_{\bl_n}(\bx_n)|<C(g,\bl_n)
	\exp (-2 \pi  g(\bx_n,\brho_n)-(2\pi-\ve) m_n(\bx))\eeq
	for any $\bl_n\in\R^n$ such that $\l_i\not=\l_j$.
\end{proposition}
Here the factor $\hat{\Delta}(\bl_n)^{N(g)}$ from Theorem \ref{theorem5.1} is hidden in the constant $C(g, \bl_n)$. Since $\Psi_{\bl_n}(\bx_n)$ is symmetric in $\bm{x}_n$, in the region 
\beq\label{g36a} x_1<x_2<\ldots< x_n\eeq
its asymptotics is described by the function
\beq\label{g36b} {\Psi'}^{as}_{\bl_n}(\bx_n)= e^{2 \pi  g(\bx_n,\brho_n)}\sum_{\sigma\in S_n}e^{2\pi \imath \sum_{i = 1}^n x_i\l_{\sigma(i)}} \prod_{{i<j}}\ccm^{-1}(\l_{\sigma(j)}-\l_{\sigma(i)})\eeq
so that for any $\bl_n\in\R^n$ (with $\l_i\not=\l_j$ for $i\not=j$) and $\bx_n$ in the region \rf{g36a}
\beq\label{g36c} |\Psi_{\bl_n}(\bx_n)-{\Psi'}^{as}_{\bl_n}(\bx_n)|<C(g,\bl_n)
\exp (2 \pi  g(\bx_n,\brho_n)-(2\pi-\ve) m'_n(\bx)),\eeq
where 
\beqq m'_n = \min_i(x_{i+1}-x_i). \eeqq

On the other hand, the asymptotics $F^{as}_{\bl_n}(\bx_n)$ of the Heckman--Opdam $A_{n-1}$ hypergeometric function $F_{\bl_n}(\bx_n)$ (extended to $\mathfrak{gl}_n$) in the region \rf{g36a} is given by \cite{HO,O}
\beq\label{g36d}F^{as}_{\bl_n}(\bx_n)= \a_n(g) \, e^{2 \pi  g(\bx_n,\brho_n)}\sum_{\sigma\in S_n}e^{2\pi \imath \sum_{i = 1}^n x_i\l_{\sigma(i)}} \prod_{{i<j}}\frac{\Gamma\bigl(\imath(\l_{\sigma(j)}-\l_{\sigma(i)})\bigr)}{\Gamma\bigl(\imath(\l_{\sigma(j)}-\l_{\sigma(i)})+g\bigr)}, \eeq
where 
\beq\label{g36e}\a_n(g) = { \prod_{k=1}^n \frac{\Gamma(kg)}{\Gamma(g)}}.
\eeq
Comparing \rf{g36b} and \rf{g36d} (recall that $\hat{w}^{-1}(\lambda) = \Gamma(\imath \l) \Gamma(-\imath \l + g)$), we see that
\beq\label{g36f}F^{as}_{\bl_n}(\bx_n)=\a_n\prod_{i\not=j}\Gamma^{-1}(\imath(\l_i-\l_j)+g) \, {\Psi'}^{as}_{\bl_n}(\bx_n).\eeq
Thus, the functions 
\beqq F_{\bl_n}(\bx_n) \qquad \text{and}\qquad \a_n\prod_{i\not=j}\Gamma^{-1}(\imath(\l_i-\l_j)+g) \, {\Psi}_{\bl_n}(\bx_n)\eeqq
diagonalize the same Calogero--Sutherland differential operators, are analytical in a neighbourhood of the origin (see~\cite[Theorem 6.1]{HR1})  and have the same asymptotics in the region \rf{g36a}. Then the uniqueness theorem \cite{HO} says that they coincide.
\begin{theorem}\label{theorem5.4}
\beq\label{g36g}
F_{\bl_n}(\bx_n)= \a_n\prod_{i\not=j}\Gamma^{-1}(\imath(\l_i-\l_j)+g) \, {\Psi}_{\bl_n}(\bx_n).\eeq
\end{theorem}

In Appendix \ref{Norm} we also prove the following proposition by calculating the Mellin--Barnes integral~\eqref{g11} at the zero point, which verifies the normalization condition \eqref{norm}.

\begin{proposition} For $\bl_n \in \mathbb{R}^n$ we have
	\begin{align}\label{Psi0}
		\Psi_{\bm{\lambda}_n}(0, \dots, 0) = \prod_{k=1}^n \frac{\Gamma(g)}{\Gamma(kg)}\, \prod_{1 \leq i \not= j \leq n} \, \Gamma(\imath( \lambda_{i} - \lambda_j) + g).
	\end{align}
\end{proposition}

The conjugation by the product $\prod_{i\not=j}\Gamma^{-1}(\imath(\l_i-\l_j)+g)$ transforms dual Hamiltonians~\rf{g1}
\begin{align}
	\ccM_r=
	(-1)^{r(n-1)}\sum_{\substack{I\subset[n] \\ |I|=r}}
	\prod_{\substack{i\in I \\ j\notin I}}
	\frac{\l_{i}-\l_{j}  - \i (1 - g)}{\l_{i}-\l_{j}}
	\prod_{i\in I}T_{\l_{i}}
\end{align}
 to the same operators with reflected coupling constant (and without sign factor) 
\beq\label{g36h} \hat{H}_r=\sum_{\substack{I\subset[n] \\ |I|=r}}
\prod_{\substack{i\in I \\ j\notin I}}
\frac{\l_{i}-\l_{j}  - \i g}{\l_{i}-\l_{j}}
\prod_{i\in I}T_{\l_{i}},
\eeq
that is
\beq \label{MH-sim} \hat{M}_r \, \prod_{i\not=j}\Gamma(\imath(\l_i-\l_j)+g)=\prod_{i\not=j}\Gamma(\imath(\l_i-\l_j)+g) \, \hat{H}_r. \eeq
This is in accordance with the known fact that Heckman--Opdam function satisfies difference equations \cite{C, DE}
\beq\label{g36k}
\hat{H}_r{F}_{\bl_n}(\bx_n)=e_r(e^{2\pi x_1},\ldots, e^{2\pi x_n}){F}_{\bl_n}(\bx_n).
\eeq
As a direct corollary of the relation \rf{g36g}, we have two integral representations for Heckman--Opdam hypergeometric function
\begin{align*}\nonumber 
	{F}_{\bl_n}(\bx_n)& = \a_n\prod_{i\not=j}\Gamma^{-1}(\imath(\l_i-\l_j)+g)\ \cL_{n}(\l_n) \, 
	\cdots \cL_{2}(\l_2) \, e^{\const \l_1x_1} \\
& =\a_n\prod_{i\not=j}\Gamma^{-1}(\imath(\l_i-\l_j)+g)\ \ccL_n(x_n)\,
\cdots\,
 {\ccL}(x_2)\,e^{2\pi i \l_1x_1}.
\end{align*}

Conversely, the Mellin--Barnes integrals are standardly converted into series by residue technique. For example, in the simplest case $n = 2$ we have
\begin{multline}
	\Psi_{\lambda_1, \lambda_2}(x_1, x_2) = \frac{1}{2\pi \Gamma(g)} \int_{\mathbb{R}} d\gamma \; e^{2\pi \imath x_2 (\l_1 + \l_2) + 2\pi \imath (x_1 - x_2) \gamma} \\[6pt]
	\times \prod_{j = 1,2} \Gamma \biggl( \imath (\l_j - \gamma) + \frac{g}{2} \biggr) \, \Gamma \biggl( \imath (\gamma - \l_j ) + \frac{g}{2} \biggr),
\end{multline}
which for $x_1 > x_2$ can be evaluated by encircling the contour in the upper half-plane. Two series of the integrand poles 
\begin{align}
	\gamma = \lambda_j + \imath \frac{g}{2} + \imath m, \qquad m \in \mathbb{Z}_{\geq 0 }, \qquad j = 1, 2
\end{align}
lead to the splitting of the wave function
\begin{align} \label{Psi-HC-2}
	\Psi_{\lambda_1, \lambda_2}(x_1,x_2) = \psi_{\lambda_1, \lambda_2}(x_1, x_2) + \psi_{\lambda_2, \lambda_1}(x_1, x_2)
\end{align}
into two series of hypergeometric type
\begin{multline} \label{HC-2}
	\psi_{\lambda_1,\lambda_2}(x_1, x_2) = e^{2\pi \bigl( \imath x_1 \lambda_1 + \imath x_2 \lambda_2 - \frac{g}{2}(x_1 - x_2) \bigr)} \sum_{m = 0}^\infty \frac{(-1)^m \, \Gamma(g + m)}{m! \, \Gamma(g)} \\[6pt]
	\times \Gamma(\imath (\lambda_1 - \lambda_2) - m) \, \Gamma(\imath (\lambda_2 - \lambda_1) + m + g) \, e^{2\pi m (x_2 - x_1)}.
\end{multline}
As explained in~\cite[Section 4]{KK1}, the formula~\eqref{Psi-HC-2} is equivalent to the relation between associated Legendre functions of the first and the second kinds~\cite[\href{http://dlmf.nist.gov/14.9.E12}{(14.9.12)}]{DLMF}
\begin{align}
	\frac{e^{-\pi \imath \mu}}{\cos(\pi \nu)} \bigl( Q_{-\nu - 1}^\mu(z) - Q_\nu^\mu(z) \bigr) = \Gamma(\mu + \nu + 1) \, \Gamma(\mu - \nu) \, P_{\nu}^{-\mu}(z).
\end{align}
Below we describe its generalization to the case $n > 2$.

Let $M_n$ be the set of lower triangular matrices with zeros on diagonal, whose elements are nonnegative intergers 
\beq\label{g38} M_n=\bigl\{M = (m_{ij})_{i,j=1}^n \, \big| \, m_{ij}\in\Z_{\geq 0}, \; j<i; \; m_{ij} = 0, \; i \leq j \bigr\}.\eeq
For a given $M=(m_{ij})\in M_n$ define the meromorphic function
\begin{align}\label{g38a} & c_M(\bl_n)=\prod_{1\leq j<i\leq n}\frac{(-1)^{m_{ij}}\Gamma(g+m_{ij})}{m_{ij}!\Gamma(g)} \; \frac{\prod\limits_{k=3}^n\prod\limits_{s=1}^{k-1}\prod\limits_{\stackrel{t=1}{t\not=s}}^{k-1} \ccm( \l_{s}-\l_{t}+\imath(p_{ks}-p_{k,t}))}{\prod\limits_{k=2}^n\prod\limits_{s=1}^{k-1}\prod\limits_{\stackrel{t=1}{t\not=s}}^k \ccm( \l_{s}-\l_{t}+\imath(p_{ks}-p_{k+1,t}))}=\\  \nonumber
&\!\! \prod_{1\leq j<i\leq n}\!\!\! \frac{(-1)^{m_{ij}}\Gamma(g+m_{ij})}{m_{ij}!\Gamma(g)} \;
\frac{\prod\limits_{k=2}^n\prod\limits_{s=1}^{k-1}\prod\limits_{\stackrel{t=1}{t\not=s}}^k \!\Gamma(\imath( \l_{s}-\!\l_{t})-\!p_{ks}+\!p_{k+1,t})\Gamma(\imath( \l_{t}-\!\l_{s})+\!p_{ks}-\!p_{k+1,t}+\!g)}{\prod\limits_{k=3}^n\prod\limits_{s=1}^{k-1}\prod\limits_{\stackrel{t=1}{t\not=s}}^{k-1} \Gamma( \imath(\l_{s}-\l_{t})-p_{ks}+p_{k,t})\Gamma( \imath(\l_{t}-\l_{s})+p_{ks}-p_{k,t}+\!g)}.
\end{align}
Here
\beq\label{g39} p_{i,j}=m_{i,j}+m_{i+1,j}+m_{i+2,j}+\ldots+m_{n,j}, \qquad p_{n + 1, j} = 0.\eeq
Let $\psi_{\bl_n}(\bx_n)$ be the  series
 \beq\label{g40} \psi_{\bl_n}(\bx_n)=e^{2\pi\left( \sum\limits_{i=1}^n\imath x_i\l_i-g(\bx_n,\brho_n)\right)}\sum\limits_{M\in M_n} c_M(\bl_n)\ e^{2\pi \sum\limits_{j<i}m_{ij}(x_i-x_j)}.\eeq 
This is the Harish-Chandra type series generalizing the two-variable functions~\eqref{HC-2}.
The residue calculation shows that the wave function $\Psi_{\bl_n}(\bx_n)$ is the symmetrization over spectral variables of the function~$\psi_{\bl_n}(\bx_n)$. 
\begin{proposition}\label{proposition5.3}
	In the region $x_1 > x_2 > \ldots > x_n$ we have
	\beq\label{g41} \Psi_{\l_1,\ldots, \l_n}(\bx_n)=\sum_{\sigma\in S_n} \psi_{\l_{\sigma(1)},\ldots, \l_{\sigma(n)}}(\bx_n).
	\eeq
\end{proposition}

\begin{proof}
	The proof is by induction. The base case $n = 1$ is trivial. For $n > 1$ we evaluate the Mellin--Barnes representation~\eqref{g11} 
	\begin{align}
		\Psi_{\bl_n}(\bx_n) =  \hat{d}_{n - 1}(g) \int_{\mathbb{R}^{n - 1}} d\bg_{n - 1} \; \hat{\Lambda}_{n - 1}(\bl_n, \bg_{n - 1}; x_n) \, \Psi_{\bg_{n - 1}} (\bx_{n - 1})
	\end{align}
	by residue technique. The integrand has poles in the upper half-plane at the points
	\begin{align} \label{g-l-poles}
		\gamma_j = \lambda_{\sigma(j)} + \imath \frac{g}{2} + \imath m_j, \qquad m_j \in \mathbb{Z}_{\geq 0}, \qquad \sigma \in S_n.
	\end{align}
	Notice that there are no poles with one $\lambda_i$ for two integration variables $\gamma_j$ and $\gamma_{k}$, since the measure function
	\begin{align}
		\hat{\mu}(\bg_{n - 1}) = \prod_{j\neq k} \Gamma^{-1}(\imath(\gamma_j - \gamma_k)) \,  \Gamma^{-1}(\imath(\gamma_j - \gamma_k) + g)
	\end{align}
	 has zeros at the points $\gamma_j - \gamma_k \in \imath\mathbb{Z}$. Besides, by Theorem~\ref{theorem5.4} the function 
	 $$\prod_{j \neq k} \Gamma^{-1}(\imath(\gamma_j - \gamma_k) + g) \, \Psi_{\bg_{n - 1}}(\bx_{n - 1})$$ 
	 modulo constant coincides with the Heckman--Opdam function $F_{\bg_{n - 1}}(\bx_{n - 1})$, which is known to be entire in spectral parameters~\cite[Theorem 2.8]{O2}.
	
	Summing up residues at the poles~\eqref{g-l-poles}, using induction assumption and symmetry of the kernel $ \hat{\Lambda}_{n - 1}(\bl_n, \bg_{n - 1}; x_n)$ with respect to $\gamma_j$ we obtain the expansion
	\begin{align}
		\Psi_{\bl_n}(\bx_n) =  \sum_{\sigma\in S_n} \tilde{\psi}_{\l_{\sigma(1)},\ldots, \l_{\sigma(n)}}(\bx_n)
	\end{align}
	in terms of the series
	\begin{align} \label{psi-tild}
		\tilde{\psi}_{\bl_n}(\bx_n) = \sum_{m_1, \dots, m_{n - 1} = 0}^{\infty} A_{\bm{m}_{n - 1}}(\bl_n; x_n) \, \psi_{\bl_{n - 1} + \imath \frac{g}{2} \bm{e}_{n - 1} + \imath \bm{m}_{n - 1}}(\bm{x}_{n - 1}),
	\end{align}
	where $\bm{e}_{n - 1} = (1, \dots, 1)$ and the coefficients 
	\begin{align}
		A_{\bm{m}_{n - 1}}(\bl_n; x_n) = (n-1)! \, \hat{d}_{n - 1}(g)  \Biggl[ \prod_{j = 1}^{n - 1} 2\pi \imath \Res_{\gamma_j = \lambda_j + \imath \frac{g}{2} + \imath m_j} \Biggr] \,  \ccL(\bl_n, \bg_{n - 1}; x_n).
	\end{align}
	Explicitly, they are given by
	\begin{multline}
		A_{\bm{m}_{n - 1}}(\bl_n) =  e^{2\pi \imath x_n \bigl(\lambda_n - \imath g \frac{n - 1}{2} - \imath \underline{\bm{m}}_{n - 1} \bigr)}  \, \prod_{j = 1}^{n - 1} \frac{(-1)^{m_j} \, \Gamma(g + m_j)}{m_j! \, \Gamma(g)} \\
		\times \frac{\prod\limits_{s = 1}^{n - 1} \prod\limits_{\substack{t = 1 \\ t\neq s}}^n \Gamma(\imath (\lambda_s - \lambda_t) - m_s) \, \Gamma(\imath(\lambda_t - \lambda_s) + m_s + g) }{\prod\limits_{1 \leq s \neq t \leq n - 1} \Gamma(\imath (\lambda_s - \lambda_t) - m_s + m_t) \, \Gamma(\imath(\lambda_t - \lambda_s) + m_s - m_t + g) }.
	\end{multline}
	It is straightforward to check that $\tilde{\psi}_{\bl_n}(\bx_n) = \psi_{\bl_n}(\bx_n)$. 
	
	Standard bounds on gamma function $\Gamma(\l)$ (Stirling formula in the cone $|\arg \l|<\pi$ together with reflection formula between the poles $\l \in \mathbb{Z}_{\leq 0}$) show that the above calculation is justified in the region $x_1 > x_2 > \ldots > x_n$ with $m_n(x)>\delta$ for some fixed $\delta>0$. 
\end{proof}

Correspondingly, Heckman--Opdam function $F_{\bl_n}(\bx_n)$ in the region $x_1 > \ldots > x_n$ is 
\beq  \label{g42} F_{\bl_n}(\bx_n)=\a_n\prod_{i\not=j}\Gamma^{-1}(\imath(\l_i-\l_j)+g)\sum_{\sigma\in S_n}\psi_{\l_{\sigma(1)},\ldots, \l_{\sigma(n)}}(\bx_n).\eeq
The formula~\eqref{g40} can be regarded as a rational counterpart of the Noumi--Shiraishi series \cite{NS}.

	\section*{Acknowledgments}
	We are immensely grateful to S. Kharchev for collaboration on the previous stage of the present work and M. Halln\"as for useful discussions. S. D. and S. Kh. thank BIMSA for hospitality. A big part of the work was done during their visit to BIMSA. Section 3 of this work was done within a research project implemented as part of the Basic Research Program at the HSE University. Section 4 was written under support by Russian Science Foundation, project No. 23-11-00150. Section 2 and Appendix B were written with the support of the Grant from the President of the Russian Federation for students of Master's degree programs.

	\section*{Appendix}
	\appendix
	\section{Double sine function and its limits} \label{AppA}
	\subsection{Double sine function} \label{AppA1}
	The  double sine  function $S_2(z):=S_2(z|\bo)$, see \cite{Ku} and references therein, is a meromorphic function that satisfies two functional relations
	\beq\label{trig3}  \frac{S_2(z)}{S_2(z+\o_1)}=2\sin \frac{\pi z}{\o_2},\qquad \frac{S_2(z)}{S_2(z+\o_2)}=2\sin \frac{\pi z}{\o_1}
	\eeq
	and inversion relation
	\beq \label{S2-sin} S_2(z)S_2(-z)=-4\sin\frac{\pi z}{\o_1}\sin\frac{\pi z}{\o_2},\eeq
	or equivalently
	\beq\label{S2-refl} S_2(z)S_2(\o_1+\o_2-z)=1. \eeq
	The function $S_2(z)$ has poles at the points
	\beq \label{S-poles}
	z = \o_1 + \o_2 + m \o_1 + k\o_2, \qquad m,k \in \mathbb{Z}_{\geq 0}
	\eeq
	and zeros at
	\beq\label{S-zeros}
	z=-m\o_1-k\o_2,\qquad m,k \in \mathbb{Z}_{\geq 0}.
	\eeq
	For $\o_1 / \o_2 \not\in \mathbb{Q}$ all poles and zeros are simple.
	In the region of analyticity $ \Re z \in ( 0, \Re(\omega_1 + \omega_2) )$ we have the following integral representation for the logarithm of $S_2(z)$
	\begin{align}\label{S2-int}
		\log S_2(z) = \frac{\pi \imath}{2} B_{2,2}(z) + \int_{\mathbb{R} + \imath 0} \frac{e^{zt}}{(e^{\omega_1 t} - 1)(e^{\omega_2 t} - 1)}\frac{dt}{t},
	\end{align}
	where $B_{2,2}(z)$ is the second order multiple Bernoulli polynomial
	\begin{align}
		B_{2,2}(z) = \frac{1}{\omega_1 \omega_2} \biggl( \Bigl( z - \frac{\omega_1 + \omega_2}{2} \Bigr)^2 - \frac{\omega_1^2 + \omega_2^2}{12} \biggr).
	\end{align}
	It is clear from this representation that the double sine function is homogeneous
	\beq\label{S-hom}
	S_2( \gamma z | \gamma\o_1, \gamma \o_2 ) = S_2(z|\o_1, \o_2), \qquad \gamma \in (0, \infty)
	\eeq
	and invariant under permutation of periods
	\beq\label{A6}
	S_2(z| \o_1, \o_2) = S_2(z | \o_2, \o_1).
	\eeq
	The double sine function can be expressed through the Barnes double gamma function $\Gamma_2(z|\bo)$ \cite{B},
	\beq
	S_2(z|\bo)=\Gamma_2(\o_1+\o_2-z|\bo)\Gamma_2^{-1}(z|\bo),
	\eeq
	and its properties follow from the corresponding properties of the double gamma function.
	
	In the limit the double sine function reduces to the usual gamma function
	\begin{align} \label{Sgamma-lim}
		S_2(z | \hat{\bo}  ) \sim \sqrt{2\pi} \, \biggl( \frac{\omega_1}{2\pi \omega_2} \biggr)^{\omega_1 z - \frac{1}{2}} \, \Gamma^{-1}( \omega_1 z), \qquad \omega_2 \to 0^+,
	\end{align}
	where $\hat{\bo} = (\omega_1^{-1}, \omega_2^{-1})$. Hence, we can compute nonrelativistic limits
	of the dual weight and kernel functions from Section~\ref{sec:hyp-ruij}
	\begin{align}
		& \rrm(\l)=S_2(\imath \l | \hat{\bo} ) \, S_2 \Big( -\imath \l+ \frac{\rg}{\omega_1 \omega_2} \, \Big| \hat{\bo} \Big), \\[6pt] 
		& \rrK(\l) =S_2^{-1}\Bigl(\imath \l +\frac{\rg}{2\omega_1 \omega_2}\Big| \hat{\bo} \Bigr) \, S_2^{-1}\Bigl(-\imath \l+\frac{\rg}{2\omega_1 \omega_2}\Big| \hat{\bo} \Bigr).
	\end{align}
	Namely, setting $\o_1 = 1$ and $\rg = g \o_2$ we have
	\begin{align} \label{Kw-dual-lim}
		\begin{aligned}
			& \hat{w}^R(\l) \sim \frac{2\pi}{(2\pi \o_2)^{g - 1}} \, \Gamma^{-1}(\imath \l) \, \Gamma^{-1}(-\imath \l + g), \\[6pt]
			& \hat{K}^R(\l) \sim \frac{(2\pi \o_2)^{g - 1}}{2\pi} \, \Gamma \Bigl( \imath \l + \frac{g}{2} \Bigr) \, \Gamma \Bigl( -\imath \l + \frac{g}{2} \Bigr),
		\end{aligned} \qquad \o_2 \to 0^+.
	\end{align}
	
\subsection{Uniform limits of kernel and measure functions} \label{AppB}
In this section we fix the period $\o_1 = 1$ and denote by $S_2(z)$ the function 
 \beq\label{init0} S_2(z):= S_2(x|1,\o_2).\eeq
 Set
 \begin{equation} \label{Pi123}
 	\begin{split}
 		\Pi_1(\epsilon) =& \{z \colon ~\epsilon \leq \Re(z) \leq 1-\epsilon\},\\[6pt]
 		\Pi_2(\epsilon) =& \{z \in \iota\mathbb{R} \colon~|\Im(z)| \geq \epsilon\}, \\[6pt] \Pi_3(\epsilon) = &\{z \in \iota\mathbb{R} \colon~|\Im(z)| < \epsilon\}.
 \end{split}\end{equation}
 In this section we prove the following statements, where we assume $g > 0$.
 
 \begin{proposition}\label{pOne} For any $\epsilon > 0$ there exists $a(\epsilon)$ such that
 	\begin{equation}
 		\left|\frac{S_2(z)}{S_2(z+g\omega_2)}\right| = \big|\big(2\sin(\pi z)\big)^g\big|(1+O(\omega_2)), ~~ z\in\Pi_1(\epsilon), ~~ 0\leq \omega_2 \leq a(\epsilon).
 \end{equation}\end{proposition}
 
 \begin{proposition}\label{pTwo} For any $\epsilon > 0$ there exists $a(\epsilon)$ such that
 	\begin{equation}
 		\left|\frac{S_2(z)}{S_2(z+g\omega_2)}\right| = \big|\big(2\sin(\pi z)\big)^g\big|(1+O(\omega_2)), ~~ z\in \Pi_2(\epsilon), ~~ 0\leq \omega_2 \leq a(\epsilon).
 \end{equation}\end{proposition}
 
 \begin{proposition}\label{pThree} For any $\epsilon > 0$ there exists $a(\epsilon)$ such that
 	\begin{equation}\label{V3}
 		\left|\frac{S_2(z)}{S_2(z+g\omega_2)}\right| \leq C(\epsilon), ~~ z\in \Pi_3(\epsilon), ~~ 0\leq \omega_2 \leq a(\epsilon).
 \end{equation}\end{proposition} 
  Propositions \ref{proposition3.1} and \ref{proposition3.2}, constituting  the base for limiting procedure of Section \ref{section3.1}, are direct corollaries of Propositions \ref{pOne}--\ref{pThree}. Indeed,  due to Proposition \ref{pOne},
   \beq \label{V1}
   	\left|\rK(x | 1, \omega_2, g\omega_2)\right|=\left|\frac{S_2(\imath x+\frac{1+\o_2}{2}+\frac{g\o_2}{2})}{S_2(\imath x+\frac{1+\o_2}{2}-\frac{g\o_2}{2})}\right| = \left|2\ch(\pi x)\right|^{-g}\left(1+O(\o_2)\right)\!.
   \eeq
For $x, g \in \mathbb{R}$ both kernel functions in the LHS and in the RHS of \rf{V1} are positive real functions, therefore \rf{V1} implies
 pointwise limit \rf{s14}. Analogously, for $x\not=0$
 \beq \label{V2} \begin{split} &\left|\rmu(x | 1, \omega_2, g\omega_2)\right|= \left| \frac{S_2(\imath x)}{S_2(\imath x+g\o_2)} \frac{S_2(-\imath x)}{S_2(-\imath x+g\o_2)}   \right|=\left|4\sh^2 (\pi x)\right|^g\left(1+O(\o_2)\right),\end{split}\eeq
and both measure functions in  the LHS and in the RHS of \rf{V2} are positive real functions, therefore \rf{V2} implies pointwise limit \rf{s15}. Besides, both sides vanish at $x=0$ (assuming $g > 0$).
 
 The bound  \rf{V1} combined with the inequality 
 $$ |\ch^{-g} \pi x|<C_1e^{-g|x|}$$
 gives the bound \rf{s16} uniform for $x \in \mathbb{R}$.
 The bound \rf{V2} together with the inequality 
$$ |\sh^{g} \pi x|<C_2e^{g|x|},\qquad |x|>\ve,$$
 gives the bound \rf{s17} uniform for $x \in \mathbb{R}$ such that $|x|>\ve$ with some fixed $\ve>0$,
 and the inequality \rf{V3} of Proposition \ref{pThree} allows to extend the bound \rf{s17} uniformly for all $x \in \mathbb{R}$.

Here is the plan of the proofs of Propositions \ref{pOne}--\ref{pThree}. We use the integral presentation~\eqref{S2-int}
\begin{equation}\begin{split}\label{dbl_sine}
\log S_2(z) & = \frac{\pi\iota}{2\omega_2}  (z^2-(1+\omega_2)z + \omega_2) + \int_{\mathbb{R}+\iota0}\frac{e^{zt}}{(e^{t}-1)(e^{\omega_2t}-1)}\frac{dt}{t},
\end{split}\end{equation}
which holds for $0 < \Re(z) < 1 + \omega_2$, 
combined with similar representation of the logarithm of sine function
\begin{equation}\label{sine}
\log\big(2\sin(\pi z)\big) = \frac{\pi \iota}{2} - \pi \iota z - \int_{\mathbb{R}+\iota0}\frac{e^{zt}}{e^{t}-1}\frac{dt}{t}, \qquad 0 < \Re(z) < 1.
\end{equation}
Direct usage of (\ref{dbl_sine}) and (\ref{sine}) gives
 \begin{align} \label{dsine} \log \frac{S_2(z+g\o_2)}{S_2(z)}=
	\frac{\pi \imath g}{2}\left({2z}+{(g-1
		)\o_2}-1\right)+\!\!\int\limits_{\R+\imath 0}\!\!\frac{e^{zt}(e^{g\o_2t}-1)}{(e^{t}-1)(e^{\o_2t}-1)}\frac{dt}{t},
\end{align}
and
\begin{equation}\begin{split}\label{closed}
& \log\frac{S_2(z)}{S_2(z+g\omega_2)}-g \log\big(2\sin(\pi z)\big) = -\frac{\pi\iota\omega_2}{2} g(g-1)-  \\[6pt]
& \int_{\mathbb{R}+\iota0}\frac{e^{zt}}{e^{t}-1}\left(\frac{e^{g\omega_2t}-1}{e^{\omega_2t}-1}-g\right)\frac{dt}{t} = - \, v.p.\int_{\mathbb{R}}\frac{e^{zt}}{e^{t}-1}\left(\frac{e^{g\omega_2t}-1}{e^{\omega_2t}-1}-g\right)\frac{dt}{t}
\end{split}\end{equation}
for $0 < \Re(z) < 1 + (1-g)\omega_2$.

Note that the real parts of equalities \rf{sine}, \rf{dsine}
and thus of \rf{closed} take place in a wider regions, including the line $\Re z=0$, $z\not=0$:
\begin{align}\label{dsine2}
	& \log\big|2\sin(\pi z)\big| =  \pi \Im z - \Re \int\limits_{\mathbb{R}+\iota0}\frac{e^{zt}}{e^{t}-1}\frac{dt}{t},\, \quad 0 \leq \Re(z) < 1,\quad z\not=0, \\[6pt] \label{dsine3} 
	& \log \biggl| \frac{S_2(z+g\o_2)}{S_2(z)} \biggr|=
	-\pi g\Im z+\Re\int_{\R+\imath 0}\frac{e^{zt}(e^{g\o_2t}-1)}{(e^{t}-1)(e^{\o_2t}-1)}\frac{dt}{t},\\[6pt] \notag
	& \hspace{4cm} 0 \leq \Re(z) < 1 + (1-g)\omega_2,\quad z\not=0.
\end{align}
Indeed, the left hand sides of \rf{sine}, \rf{dsine} have limits as $z\to \imath y$, $y\in\R \setminus \{0\}$ due to continuity. Thus, the right hand sides have the same limits as well. It remains to verify that two limits in the definition of limiting integrals: $z\to \imath y$ and $N\to \infty$, can be interchanged. The latter designates the border of finite segment in the definition of improper integral. But the integrals on infinite segments $[N,\infty)$ and $(-\infty,-N]$ can be uniformly bounded for $N$ big enough by arbitrary small quantity once $y\not=0$ is fixed. 

 The uniform bound of the integral on $[N,\infty)$ is extracted from the absolute convergence of the integrals in considerations on the edge $t\to +\infty$. On the edge $t\to -\infty$ we have oscillating integrals of the type $\int e^{iyt}f(t)$, and the oscillation period depends only on~$y$. We then can bound the oscillating integral by the integral over one period in a region when the function $f(t)$ monotonically decrease.

The bounds for the above three integrals are covered by the following five technical lemmas, which are proved in Appendix~\ref{AppA3}. The first two of them give us general bounds for $\Re z \in [0, 1-\epsilon)$ and correspond to the interval $(-1,+\infty)$.

Denote 
\begin{equation}
\Omega = \frac{-e^{zt}}{e^{t}-1}\left(\frac{e^{g\omega_2t}-1}{e^{\omega_2t}-1}-g\right)\frac{dt}{t}.
\end{equation}

\begin{lemma}\label{lOne} For any $\epsilon > 0$ there exists $a(\epsilon)$ such that for $0\leq \omega_2\leq a(\epsilon)$, $\Re z \in [0, 1-\epsilon)$
\begin{equation}
\left|\int^{+\infty}_{1}\Omega\right| 
\leq \omega_2 C_1,
\end{equation}
where $C_1$ does not depend on either $z$ or $\omega_2$.\end{lemma}

\begin{lemma}\label{lTwo} For any $\epsilon > 0$ there exists $a(\epsilon)$ such that for $0\leq \omega_2\leq a(\epsilon)$, $\Re z \in [0, 1-\epsilon)$
\begin{equation}\begin{split}
\left|\Re\left(v.p.\int^1_{-1}\Omega\right)\right|
\leq \omega_2C_2,
\end{split}\end{equation}
where $C_2$ does not depend on either $z$ or $\omega_2$.\end{lemma}

Other lemmas give bounds for the integral over $(-\infty, -1)$ for three distinguished cases, when $z$ belongs to the sets $\Pi_1$, $\Pi_2$, $\Pi_3$, see~\eqref{Pi123}.

\begin{lemma}\label{lThree} For any $\epsilon > 0$ there exists $a(\epsilon)$ such that for $0\leq \omega_2\leq a(\epsilon)$, $z \in \Pi_1(\epsilon)$ 
\begin{equation}
\left|\int^{-1}_{-\infty}\Omega\right|  
\leq \omega_2 C_3,
\end{equation}
where $C_3$ does not depend on either $z$ or $\omega_2$.\end{lemma}

\begin{lemma}\label{lFour} Let $g\geq 1$. Then for any $\epsilon > 0$ there exists $a(\epsilon)$ such that for  $0\leq \omega_2 \leq a(\epsilon)$, $z\in\Pi_2(\epsilon)$
\begin{equation}
\left|\int^{-1}_{-\infty}\Omega\right| \leq \frac{\omega_2}{|z|}C_4,
\end{equation}
where $C_4$ does not depend on either $z$ or $\omega_2$.\end{lemma}

\begin{lemma}\label{lFive} For any $\epsilon > 0$ there exists $a(\epsilon)$ such that for $0\leq \omega_2\leq a(\epsilon)$, $z\in\Pi_3(\epsilon)\backslash\{0\}$ 
\begin{equation}
\Re\left(\int^{-1}_{-\infty}\Omega\right) \leq C_5  + (1-g)\log|z|,
\end{equation}
where $C_5$ does not depend on either $z$ or $\omega_2$.
\end{lemma}

\begin{remark} Note that on the left hand side of (\ref{closed}) logarithm of sine grows with real asymptotics $-g\log|z|$, $z\in\Pi_3\backslash\{0\}$. So it cancels bound $-g\log|z|$ from Lemma \ref{lFive}, when we move back to the ratio of double sines.
\end{remark}

{\bf Proofs of Propositions \ref{pOne}--\ref{pThree}.} Using Lemmas \ref{lOne},~\ref{lTwo} and~\ref{lThree} we obtain 
\begin{equation*}\begin{split}
\left|\Re\left(\log\frac{S_2(z)}{S_2(z+g\omega_2)}-g \log\big(2\sin(\pi z)\big)\right)\right| \leq \omega_2(C_1+C_2+C_3),\\
0\leq \omega_2\leq a(\epsilon), ~~ z\in\Pi_1(\epsilon).
\end{split}\end{equation*}
Hence,
\begin{equation*}\begin{split}
e^{-\omega_2(C_1+C_2+C_3)} \leq \left|\frac{S_2(z)}{S_2(z+g\omega_2)}\cdot\big(2\sin(\pi z)\big)^{-g}\right| \leq e^{\omega_2(C_1+C_2+C_3)},\\
0\leq \omega_2\leq a(\epsilon), ~~ z\in\Pi_1(\epsilon),
\end{split}\end{equation*}
and Proposition \ref{pOne} is proved. 

By applying Lemma \ref{lFour} instead of Lemma \ref{lThree}, we obtain the statement of Proposition~\ref{pTwo} for $g \geq 1$.
For $0<g<1$ we then can write, using difference equation for the double sine function,
\begin{equation*}\begin{split}
	& \left|\frac{S_2(\iota x)}{S_2(\iota x+g\omega_2)}\right| =  \left|\frac{S_2(\iota x)}{S_2(\iota x+(g+1)\omega_2)}\right|\cdot \left|\frac{1}{2\sh{\pi (x-\iota g\omega_2)}}\right|\\[6pt]
	& = |2\sh \pi x|^g\left|\frac{\sh \pi x}{\sh\pi(x-\iota g\o_2)}\right|\left(1+O(\o_2)\right)= |2\sh \pi x|^g\left(1+O(\o_2)\right),
	\end{split}
\end{equation*}
because 
\beqq \left|\frac{\sh \pi x}{\sh\pi (x-\iota g\o_2)}\right|=1+O(\o_2) \qquad\text{for} \qquad |x|>\ve. \eeqq
The last inequality follows from Taylor formula since logarithmic deri\-vative of $\sh (\pi x)$ is $\pi\cth \pi x$ which is bounded on the interval $|x|>\ve$.

To prove Proposition \ref{pThree} apply Lemmas \ref{lOne},~\ref{lTwo} and~\ref{lFive} to (\ref{closed})
\begin{equation*}\begin{split}
& \Re\left(\log\frac{S_2(z)}{S_2(z+g\omega_2)}\right) \leq \Re\Big(g \log\big(2\sin(\pi z)\big)\Big)+ \\[6pt]
&  \omega_2(C_1+C_2)+C_5 + (1-g)\log|z| \leq C_0+\omega_2(C_1+C_2)+C_5 + \log|z|,\\[6pt]
& \hspace{4cm} 0\leq \omega_2\leq a(\epsilon), ~~ z\in\Pi_3(\epsilon)\backslash\{0\},
\end{split}\end{equation*}
where we also used the fact that $\Re\big(g \log\big(2\sin(\pi z)\big)\big) \leq C_0 + g\log|z|$ with $C_0$ not depending on $z$. Exponentiating the above inequality we arrive at Proposition \ref{pThree} for $z\in\Pi_3(\epsilon)\backslash\{0\}$. One notes that $S_2(z)/S_2(z + g\omega_2)$ is continuous at $z = 0$ as well, thus we cover the case of $\Pi_3$ with the same constant $C_0$. 
\hfill{$\Box$}

\subsection{Proofs of lemmas} \label{AppA3}
{\bf 1. Proof of Lemma \ref{lOne}.} Write $\Omega = -\o_2\alpha(t)\beta_g(\o_2t)dt$, where 
\begin{equation}\label{ab}
\alpha(t) = \frac{e^{zt}}{e^{t}-1}, \qquad \beta_g(t) = \frac{1}{t}\left(\frac{e^{g t}-1}{e^{t}-1}-g\right).
\end{equation} 
It is clear that
\begin{equation}\label{ri_ind}
|\alpha(t)| = \left|\frac{e^{zt}}{e^{t}-1}\right| \leq \frac{e}{e-1}\cdot\frac{e^{\Re(z)t}}{e^{t}}, \qquad t \geq 1,
\end{equation}
and $\beta_g(t)$ is a regular function of $t$, such that
\beq\label{a71}
	\beta_g(0)=\frac{g(g-1)}{2},\qquad 
	\beta_g(t)\sim \left\{
	\begin{aligned}
		& C_1\frac{e^{({\mg} -1)t}}{t}, && \quad  t\to\infty,\\[8pt]
		& \frac{1-g}{t}, && \quad t\to-\infty .
	\end{aligned}\right.
\eeq
	Here
	\beq \label{checkg}{\mg} =\max(g,1). \eeq
Hence for $t\geq 0$ we have the uniform bound
\beq\label{ri_ind2} |\beta_g(t)|<C{e^{({\mg} -1)t}}  \eeq
valid for some constant $C(g)$. 
Combining (\ref{ri_ind}) with (\ref{ri_ind2}), we deduce that
\begin{equation}\label{const_1}
\left|\int_{t\geq 1}\Omega\right| \leq \omega_2C\frac{e}{e-1}\int_{t\geq 1}e^{-t(1-\Re(z)-({\mg} -1)\omega_2)}dt \leq \omega_2 C_1.
\end{equation}
 For the last integral to be bounded we choose $a(\epsilon)$ such that $$1-\Re(z)-({\mg}-1) \omega_2 \geq \delta > 0$$ for any $0\leq \omega_2 \leq a(\ve)$ and $\Re (z) \in[0, 1 -\ve)$. Since $$1-\Re(z)-({\mg} -1)\omega_2 > \epsilon-{\mg} \omega_2 \geq \epsilon -{\mg} a,$$ we can take $a(\epsilon) \leq \epsilon/2{\mg} $.
\hfill{$\Box$}
\bigskip

\noindent {\bf 2. Proof of Lemma \ref{lTwo}.} Firstly, note that for $-1\leq t\leq 1$ we have inequality
	\beq\label{reg_bnd}
			\left|\frac{1}{e^{t}-1}-\frac{1}{t}\right| \leq C.\eeq
	Denote by $\widehat{\Omega}$ the following form
	\begin{equation}
		\widehat{\Omega} = \frac{e^{zt}dt}{t^2}\cdot\left(\frac{e^{g\omega_2 t}-1}{e^{\omega_2t}-1}-g\right)
	\end{equation}
	and consider $$v.p.\int_{[-1,1]}\Omega = v.p.\int_{[-1,1]}\widehat{\Omega} + \int_{[-1,1]}(\Omega-\widehat{\Omega}).$$ Note that due to \rf{reg_bnd}
	\begin{equation}\label{const_m}
		\left|\int_{[-1,1]}(\Omega-\widehat{\Omega})\right| \leq C\omega_2 \int_{[-1,1]}e^{\Re(z)t} |\beta_g(\o_2 t)|dt \leq \omega_2 \widehat{C}.
	\end{equation}
for some constant $\widehat{C}$. The function $\beta_g(t)$ is defined in \rf{ab}.

	It is left to estimate the remaining term with $\widehat{\Omega}$. Denote  $z = x+\iota y$. Then 
	\begin{equation}\label{a78}\Re\bigg(v.p.\int_{[-1,1]}\widehat{\Omega}\bigg) = \o_2 \; v.p.\int_{[-1,1]}\frac{\cos(yt)e^{xt}\beta_g(\o_2 t)}{t}dt.\end{equation} 
	The integral in the right hand side of \rf{a78} is of the type
	\beqq v.p.\int_{-1}^{1}\frac{\vf(t)}{t}dt \eeqq
	where $\vf(t)$ is a function regular at $t = 0$. It can be also defined as the regular integral
	\beqq \int_{-1}^{1}\frac{\vf(t)-\vf(0)}{t}dt. \eeqq
	Let us argue that it is uniformly bounded for small enough $\o_2$. Indeed, due to~\eqref{ab} we have decomposition $$\beta_g(t) = \beta_g(0) + tB(t)$$ on $[-1,1]$, where $$\sup_{t \in [-1,1]}|B(t)| \leq B.$$ Thus, for $\omega_2 \leq 1$ we have $$\beta_g(\omega_2t) = \beta_g(0) + \omega_2 tB(\omega_2t)$$ with $$\sup_{t\in[-1,1]}|B(\omega_2t)| \leq B.$$
    This gives the desired bound for Lemma \ref{lTwo}.
\hfill{$\Box$}
\bigskip

\noindent {\bf 3. Proof of Lemma \ref{lThree}.} Here we use the bound valid for arbitrary $\ve>0$ 
\beq\label{a79}|\beta_g(t)|<C(\ve)e^{\ve |t|}, \qquad t\leq 0, \eeq
which follows from \rf{a71}, and
\begin{equation}\label{le_ind}
|\alpha(t)| = \left|\frac{e^{zt}}{e^{t}-1}\right| \leq \frac{1}{1-e^{-1}}\, e^{\Re(z)t}, \qquad t \leq -1.
\end{equation}
They imply  inequality
\begin{equation}\label{const_2}
\left|\int_{t\leq -1}\Omega\right| \leq \o_2 \frac{C(\ve)}{1-e^{-1}}\int_{t\geq 1}e^{-t(\Re(z)-\o_2\ve)}dt \leq \omega_2 C_3
\end{equation}
 where by lemma assumptions $\Re (z) \geq  \ve$, $0 \leq\o_2 \leq a(\ve)$ and we choose $a(\ve)<1$. Thus, we get the uniform bound, as stated in Lemma \ref{lThree}. 
\hfill{$\Box$}
\bigskip

\noindent {\bf 4. Proof of Lemma \ref{lFour}.}  Here we exploit the following elementary bound for oscillating integrals
$$I=\int_a^\infty e^{\pi \iota x/T}f(x)dx, $$
where $f(x)$ is a positive function, monotonically decreasing to zero at $(a,+\infty)$:
$$|I|< 2\int_a^{a+T} f(x)dx<  2T f(a).$$

Consider again the function $\beta_g(x)$, see \rf{ab}.  
We claim that for $x \leq 0$ and  $g\geq 1$ this function is non-positive and  monotonically increases from $\lim_{x\to -\infty}\beta_g(x) = 0$ to $\beta_g(0) = g(g-1)/2$. This can be observed by analyzing its dependence on $g$.

Consider
\beq\begin{split}
\frac{\partial^2 \beta_g(x)}{\partial x \partial g} = \bigg(-\frac{1}{x} + \frac{e^{gx}}{e^{x}-1}\bigg)^{\prime}_x = \frac{1}{x^2} + e^{gx} \, \frac{g(e^x-1) - e^{x}}{(e^{x}-1)^2}
\end{split}\eeq
and
\beq\begin{split}
\frac{\partial^3 \beta_g(x)}{\partial x \partial g^2} & = xe^{gx} \, \frac{g(e^{x}-1) - e^{x}}{(e^{x}-1)^2} + \frac{e^{gx}}{e^{x}-1} \\
& = \frac{e^{gx}}{(e^{x}-1)^2}\, \big[(xg+1)(e^{x}-1) - xe^{x}\big].
\end{split}\eeq
Since $x\in(-\infty, 0)$, either $xg + 1 \leq 0$ and numerator is greater than zero, or $x > -g^{-1}$ and we can estimate
\beq\begin{split}
e^{x}-1 \geq x, ~~~ x < 0,
\end{split}\eeq
so that
\beq\begin{split}
(xg+1)(e^{x}-1) - xe^{x} \geq (xg+1)x - x(1+x) = x^2(g-1).
\end{split}\eeq
We thus obtained that for $g \geq 1$
\beq\begin{split}
\frac{\partial^3 \beta_g(x)}{\partial x\partial g^2} \geq 0, \qquad x < 0.
\end{split}\eeq
So $\frac{\partial^2\beta_g(x)}{\partial x\partial g}$ is monotone in $g$ for $g \geq 1$, $x < 0$.

In the case $g = 1$ we have
\beq\begin{split}
\frac{\partial^2\beta_g(x)}{\partial x\partial g}\bigg|_{g=1} = \frac{1}{x^2} - \frac{1}{4\sh^2\frac{x}{2}}\geq0
\end{split}\eeq 
since $\sh^2 x \geq x^2$. Hence, we obtain that
\beq\begin{split}
\frac{\partial^2 \beta_g(x)}{\partial x\partial g} \geq \frac{\partial^2 \beta_g(x)}{\partial x\partial g}\bigg|_{g=1} \geq 0, ~~~ g\geq 1, ~~~ x < 0.
\end{split}\eeq
Therefore, $\frac{\partial \beta_g(x)}{\partial x}$ is monotone in $g$ for $g \geq 1$ and $x < 0$.

Since $\b_1(x)$ is identically equal to zero, the derivative $\frac{\partial \b_1(x)}{\partial x}$ is also zero at any $x$,
 which finally implies that 
 $\beta_g(x)$ is monotone for positive $x$ for any $g\geq 1$.

By lemma assumption $z = \imath y \in \imath \mathbb{R}$ and $|y| \geq \ve$. Then
\begin{equation}
\int_{t\leq -1}\Omega = \int_{t\leq -1}\frac{\omega_2 e^{\iota y t}dt}{1-e^{t}}\cdot \beta_g(t\omega_2).
\end{equation}
Since $\beta_g(t\omega_2)/(1-e^{t})$ increases monotonically from zero at $-\infty$, we have bound
\begin{equation}\label{neg_bnd}
\left|\int_{t\leq -1}\Omega\right| \leq 2\omega_2\int^{-1}_{-1- \pi/|y|}\frac{\beta_g(t\omega_2) dt}{1-e^{t}} \leq \frac{\omega_2\pi g(g-1)}{|y|(1-e^{-1})} = \frac{\omega_2}{|y|}C_g.
\end{equation}
This concludes the proof of lemma. 
\hfill{$\Box$}
\bigskip

\noindent {\bf 5. Proof of Lemma \ref{lFive}.} Denote
\begin{equation}\label{Udef}
U(t) = \frac{1}{(-t)(1-e^{t})}\left(g-\frac{1-e^{g\omega_2t}}{1-e^{\omega_2t}}\right).
\end{equation}
By lemma assumption $z = \iota y \in \imath \mathbb{R}$, so that $\Re\big(\Omega\big) = \cos(yt) U(t) dt$. Without loss of generality, assume that $y > 0$.

Let us present some properties of $U(t)$ and derive the result from them. First, note that
\begin{equation}\label{comp_zero}
(g-1)(1-e^t)U(t) \geq 0, ~~~ t < 0,
\end{equation}
and
\begin{equation}\label{as_z_mi}
\lim_{t \to -\infty}(-t)(1-e^t)U(t) = g-1, \qquad \lim_{t \to 0} (1-e^t)U(t) = \omega_2\, \frac{g(g-1)}{2}.
\end{equation}
Additionally,
\begin{equation}\label{comp_g}
(g-1)(-t)(1-e^t)U(t) \leq (g-1)^2, ~~~ t < 0.
\end{equation}
Since $1/(1-e^{t})$ is positive and increases on the interval $(-\infty,0)$, we can deduce crucial monotonicity properties of $U(t)$ that will imply the desired bounds for
\beq\label{int-cosU}
\int_{t\leq -1}\cos(yt)U(t) dt.\eeq 

According to (\ref{comp_zero}) and (\ref{comp_g}), we know that $$\sup_{t\in(-\infty, -1)}|tU(t)| \leq B$$ for $B$ not depending on the choice of $\omega_2$. Also, due to (\ref{as_z_mi}) we know asymptotics of $U(t)$ at $-\infty$:
\begin{equation}
\lim_{t\to-\infty}(-t)U(t) = g-1.
\end{equation}
We are interested in one more technical statement about monotonicity at infinity. It arises from the following properties of $U(t)$. By definitions~\eqref{ab},~\eqref{Udef}
\beq\begin{split}
U(t) = \frac{(-\omega_2)\beta_g(-\omega_2t)}{1-e^t}.
\end{split}\eeq
We claim that
\begin{enumerate}
\item[(1)]  $\beta_g(-t)$ has finite number of extremums at $(-\infty, 0)$;
\item[(2)] $U(t)$ is monotone on domain $(-\infty, -m\omega^{-1}_2)$, where $m \geq 0$ can be chosen independent of $\omega_2$.
\end{enumerate}
To deduce (1), note that $$\partial_t \beta_g = \frac{P(t,e^t,e^{gt})}{Q(t,e^t,e^{gt})}, \qquad P(x_1,x_2,x_3),Q(x_1,x_2,x_3)\in\mathbb{R}[x_1,x_2,x_3].$$ 
Real zeroes of $P$ with respect to $t$ lie on compact $I \subset \mathbb{R}$, so there are only finite number of them due to analyticity.

To prove (2), use property (1) that implies that $\beta_g(-t)$ has the leftmost extremum, say, at $t = -m \leq 0$ (if there is no such point, then we can simply take $m = 0$). Due to (\ref{as_z_mi}), $$(1-g)\beta_g(-t)$$ grows at range $t \in (-\infty, -m)$. Therefore $$(1-g)\omega_2\beta_g(-\omega_2t)$$ grows at range $t \in (-\infty, -m\omega^{-1}_2)$ and the growth of $\frac{1}{1-e^t}$ implies that $(1-g)U(t)$ also does.

Using the above properties, we can estimate the integral~\eqref{int-cosU}. Let us consider two cases.

First, suppose that $-\pi y^{-1}/2 > -m\omega^{-1}_2$. Then
\begin{equation}\begin{split}\label{inf_ext}
& \int^{-m\omega^{-1}_2}_{-\infty}\cos(y t)U(t)dt
\leq \sign(g-1)\int^{-m\omega^{-1}_2}_{-m\omega^{-1}_2-\pi y^{-1}}U(t) dt \\
& \leq \sup_{t \leq -1}\big\{|tU(t)|\big\} \int^{-my\omega^{-1}_2}_{-my\omega^{-1}_2-\pi}\frac{dt}{t} \leq \sup_{t \leq -1}\big\{|tU(t)|\big\} \int^{-\pi/2}_{-3\pi/2}\frac{dt}{t} \leq \widehat{B}.
\end{split}\end{equation}
Consider the remaining part
\begin{equation}\begin{split}\label{fin_ext}
& \int^{-1}_{-m\omega^{-1}_2}\cos(yt)U(t) dt = \int^{-1}_{-m\omega^{-1}_2}\frac{\omega_2\cos(y t)dt}{1-e^{t}}(-\beta_g(-\omega_2 t)) \\
& \leq \frac{\omega_2}{1-e^{-1}}\cdot\sup_{t \leq 0} |\beta_g(-t)|\int^{-1}_{-m\omega^{-1}_2}dt \leq \frac{m}{1-e^{-1}}\cdot\sup_{t \leq 0} |\beta_g(-t)| \leq \widehat{D}.
\end{split}\end{equation}
These estimates are sufficient for the first case.

Second, suppose that $-\pi y^{-1}/2 \leq -m\omega^{-1}_2$. Then
\begin{equation}\begin{split}\label{inf_arg}
\int^{-\pi y^{-1}/2}_{-\infty}\cos(y t)U(t) dt
\leq \sign(1-g)\int^{-\pi y^{-1}/2}_{-3\pi y^{-1}/2}\cos(y t) U(t) dt \\
\leq \sup_{t \leq -1}\big\{|tU(t)|\big\}\cdot\int^{-\pi/2}_{-3\pi/2}\frac{\cos(t)dt}{t} \leq \widehat{B}.
\end{split}\end{equation}
So, we have a bound not depending on either $\omega_2$ or $y$. Consider the remaining part of integration domain for $g \geq 1$
\begin{equation}\begin{split}\label{fin_arg}
& \int^{-1}_{-\pi y^{-1}/2}\cos(y t) U(t)dt \leq \int^{-1}_{-\pi y^{-1}/2}\frac{(g-1)dt}{(-t)(1-e^{t})} = (g-1)\int^{-1}_{-\pi y^{-1}/2}\frac{dt}{-t} \\
& + (g-1)\int^{-1}_{-\pi y^{-1}/2}\frac{dt}{(-t)(e^{-t}-1)} \leq (g-1)(-\log(|t|))\Big|^{-1}_{-\pi y^{-1}/2} \\
& + (g-1)\int^{-1}_{-\infty}
\frac{dt}{(-t)(e^{-t}-1)}\\ & = (g-1)\log(\pi y^{-1}/2) + D = (1-g)\log y+ \widehat{D}
\end{split}\end{equation}
with $D$ and $\widehat{D}$ not depending on the choice of $y$. If $0 < g < 1$, then integral is simply non-positive.

Since (\ref{inf_ext}) with (\ref{fin_ext}) cover integration domain $(-\infty,-1)$, as well as (\ref{inf_arg}) with (\ref{fin_arg}), this concludes the proof.
\hfill{$\Box$}

\setcounter{equation}{0}
\section{Asymptotics of wave function} \label{AppD}
The arguments presented here are similar to those used by M. Halln\"as and S. Ruijsenaars in \cite{HR3}. We consider Mellin--Barnes integral representation~\eqref{g11} and shift the integration contour through the nearest poles. The residue term gives the asymptotics, while for the rest we find out the proper bounds.
The result of the contour shift is described by the following lemma.

\begin{lemma}\label{wf_int_pr}
Assume that $g>1$, $\bl_n, \bx_n \in \mathbb{R}^n$. Then we have the integral representation with shifted elementary contours $\mathbb{R}\mapsto\mathbb{R}+\iota(g/2+\epsilon)$ for $\epsilon\in(0,1)$
\beq\begin{split}\label{int_pres}
& \hat\Psi_{\bm{\lambda}_n}(\bx_n) = e^{2\pi\iota x_n(\lambda_1+\ldots+\lambda_n)} \sum^{n-1}_{r = 0}\hat d_{n-1-r} \int_{(\mathbb{R} + \iota\frac{g}{2} + \iota\epsilon)^{n-1-r}}d\bm{t}_{n-1-r}\times\\[6pt]
& \sum_{\substack{J \subset {\{1,\ldots, n\}}\\|J| = r}}\hat\Psi_{(t_1,\ldots,t_{n-1-r};\{\lambda_i + \iota g/2\}_{i\in J})}(\bx_{n-1}-x_n\bm{1})\prod_{\substack{j \notin J\\k\in J}}\ccK(\lambda_j - \lambda_k - \iota g/2)\times\\[3pt]
&  \prod_{\substack{j\notin J\\l=1,\ldots,n-1-r}}\!\!\!\ccK(\lambda_j-t_l)\!\! \prod_{\substack{j\neq k\\1\leq j,k\leq n-1-r}}\ccm(t_j-t_k)\prod_{\substack{j\in J\\l=1,\ldots,n-1-r}}\!\!\!\ccK^{-1}(\lambda_j-t_l + \iota g).
\end{split}\eeq
\end{lemma}
The proof is a direct residue calculation. 

\begin{remark} Asymptotical part $\hat\Psi^{as}$ corresponds to $r = n-1$ of (\ref{int_pres}), where we take it for $\hat\Psi$ from the right as well.\end{remark}

To state and prove next lemma it is convenient to multiply $\hat\Psi_{\bm{\lambda}_n}(\bx_n)$ by an exponent associated to square root of dual measure. Recall the definition of renormalized wave function $\Phi_{\bl_n}(\bx_n)$ (\ref{g32}) and introduce
%
\def\spsy{\check}
\beq\label{B103}\begin{split}
\spsy\Phi_{\bm{\lambda}_n}(\bx_n) := e^{\pi g\sum_{1\leq i\leq n-1}|x_i-x_n|} \, \Phi_{\bm{\lambda}_n}(\bx_n).
\end{split}\eeq
\begin{lemma}\label{rwf_bnd} Assume that $0 \leq \Im\lambda_1\leq\ldots\leq \Im\lambda_n < \epsilon < 1$. Then
\beqq\begin{split}
\bigl|\spsy\Phi_{\bm{\lambda}_n}(\bx_n)\bigr| \leq C(g) \, \prod_{i<j}(1+|\Re\lambda_i-\Re\lambda_j|)^{N(n)} \,  e^{-2\pi(x_1\Im\lambda_1+\ldots +x_n\Im\lambda_n)},
\end{split}\eeqq
where $N(n)$ can be taken as
\beq \label{B105} N(n) = 4^{n-2}g - \frac{(4^{n-2}-1)(g-2)}{3}.\eeq
\end{lemma}

 Let us sketch the main steps of the proof, the details are given in Appendix~\ref{AppB1}. The proof goes by induction over $n$ with trivial base.

Indeed, looking at (\ref{int_pres}), we see that all terms contain wave function with $n-1$ arguments, hence it remains to majorate integrands (say, we fix $J\subset\{1,\ldots,n\}$) of rescaled wave functions. It can be done via Stirling formula. As a result, we bound the integrand by
\begin{multline}\label{add_term}
C(g,\epsilon) \, e^{\pi g\!\!\!\!\!\sum\limits_{1\leq i\leq n-1}\!\!|x_i-x_n|} \, \!\!\prod_{\substack{j < k\\ j,k\notin J}}(1+|\Re(\lambda_j-\lambda_k)|)^{1-g} \!\!\!\!\!\!\! \prod_{\substack{j\notin J\\l=1,\ldots,n-1-r}}\!\!\!\!\!\!\!(1+|\Re\lambda_j-t_l|)^{g-1} \\
\times  \, \exp\pi\bigg[\sum_{\substack{j<k\\j,k\notin J}}|\Re(\lambda_j-\lambda_k)| + \sum_{1\leq j<k\leq n-1-r}|t_j-t_k| - \sum_{j\notin J}\sum_{l = 1}^{n - 1 - r}|\Re\lambda_j-t_l|\bigg]\\
\times \ddelta(\bm{t}_{n-1-r})^{1-g}\big|\spsy\Phi_{(\bm{t}_{n-1-r} + (\iota g/2+\iota\epsilon)\bm{1};~\{\lambda_i+\iota g/2\}_{i\in J})}(\bx_{n-1}-x_n\bm{1})\big|,
\end{multline}
 where 
\begin{align}
	\hat{\Delta}(\bm{t}_m) = \prod_{1 \leq i < j \leq m} (1 + |t_i - t_j|). 
\end{align}
Applying induction assumption combined with shift relation for $\eta\in\mathbb{C}$
\beq\begin{split}
\hat\Psi_{(\bm{\lambda}_n + \eta\bm{1})}(\bx_n) = e^{2\pi\iota\eta(x_1+\ldots+x_n)}\hat\Psi_{(\bm{\lambda}_n)}(\bx_n);\\
\spsy\Phi_{(\bm{\lambda}_n + \eta\bm{1})}(\bx_n) = e^{2\pi\iota\eta(x_1+\ldots+x_n)}\spsy\Phi_{(\bm{\lambda}_n)}(\bx_n),
\end{split}\eeq
we reduce the problem to the following generalization of \cite[Theorem A1]{HR3}, which is proven in Appendix~\ref{AppC}.
\begin{proposition}\label{abc_int}
For $A,B,C > 0$, and real $s_1,\ldots,s_m$, there exists $D(A,B,C)$ such that
\begin{multline} \label{B108}
\bigg|\int_{\mathbb{R}^{m-1}}d\bm{t}_{m-1} \; \ddelta(\bm{s}_m)^A \, \ddelta(\bm{t}_{m-1})^C \, \prod_{j = 1}^m \prod_{l = 1}^{m - 1}(1+|s_j-t_l|)^B \\
\times \exp\pi\bigg[\sum_{1\leq j<k\leq m}|s_j-s_k| + \sum_{1\leq j<k\leq m-1}|t_j-t_k| - \sum_{j = 1}^m \sum_{l=1}^{m - 1} |s_j-t_l|\bigg]\bigg|\\[6pt]
\leq D(A,B,C)\,\ddelta(\bm{s}_m)^{A + 2B + C + 1}.
\end{multline}
\end{proposition}
This proposition ends the proof of Lemma~\ref{rwf_bnd} with some power $N(n)$. To calculate its value, combine (\ref{add_term}) with induction assumption.

\begin{corollary} \label{cor:one}
For rescaled wave function $\Phi_{\bm{\lambda}_n}(\bx_n)$, see \rf{g32}, and real parameters $\bl_n$ and $\bx_n$ we have the bound
\beq\begin{split}\label{corOne}
\big|\Phi_{\bm{\lambda}_n}(\bx_n)\big| \leq C(g) \, \ddelta(\bl_n)^{N(n)} \,  e^{-\pi g\sum_{i<j}|x_i-x_j|},
\end{split}\eeq
where $N(n)$ is given by the formula \rf{B105}.
\end{corollary}
The same ideas as in proven lemma can be used to show that $\Phi^{as}_{\bl_n}(\bx_n)$ is the leading asymptotical part of $\Phi_{\bl_n}(\bx_n)$. Introduce
\beq\begin{split}
\spsy\Phi^{as}_{\bm{\lambda}_n}(\bx_n) := e^{\pi g\sum_{i <j}|x_i-x_j|} \, \Phi^{as}_{\bm{\lambda}_n}(\bx_n),
\end{split}\eeq
where $\Phi^{as}_{\bm{\lambda}_n}(\bx_n)$ is given by the relation \rf{g33}.
\begin{lemma}\label{rwf_comp}
For $\epsilon\in(0,1)$ and $\bl_n \in\mathbb{R}^n$, we have the following bound
\beq\begin{split}
\big|\spsy\Phi_{\bm{\lambda}_n}(\bx_n) - \spsy\Phi^{as}_{\bm{\lambda}_n}(\bx_n)\big| \leq C'(g,\epsilon)\,\ddelta(\bl_n)^{N(n)}\, e^{-2\pi\epsilon \, m_n(\bx_n)}
\end{split}\eeq
with $m_n(\bx_n)$ as in (\ref{gmn}).
\end{lemma}

This lemma is proven in Appendix~\ref{AppB1}. Writing $2\pi - \epsilon$ for $\epsilon \in (0, 2\pi)$ instead of $2\pi\epsilon$ for $\epsilon\in(0,1)$, one deduces the corollary below.
\begin{corollary} \label{cor:two}
For $\epsilon\in(0,2\pi)$ and $\bl_n \in\mathbb{R}^n$, we have the following bound
\beq\begin{split}\label{corTwo}
\big|\Phi_{\bm{\lambda}_n}(\bx_n) - \Phi^{as}_{\bm{\lambda}_n}(\bx_n)\big| \leq C'(g,\epsilon) \, \ddelta(\bl_n)^{N(n)}\, e^{-2\pi g(\bx_n, \bm{\rho}_n)-(2\pi-\epsilon) m_n(\bx_n)},
\end{split}\eeq
where $\bm{\rho}_n$ is given in (\ref{grho}).
\end{corollary}

 Observe that Corollaries~\ref{cor:one},~\ref{cor:two} are exactly Theorems~\ref{theorem5.2},~\ref{theorem5.1} from Section~\ref{sec:as}. 

\subsection{Proofs of lemmas} \label{AppB1}
{\bf 1. Proof of Lemma \ref{rwf_bnd}.}  The proof goes by induction over $n$ with trivial base. Consider the formula (\ref{int_pres})
\beq\begin{split}\label{psi_dec}
\hat\Psi_{\bl_n}(\bx_n) = \sum^{n-1}_{r=0}\sum_{\substack{J \subset\{1,\ldots,n\}\\|J|=r}}I_{r,J}(\bx_n;\bl_n;\bm{t}_{n-1};\epsilon),
\end{split}\eeq
where for fixed $J\in 2^{\{1,\ldots,n\}}$ with $|J| = r$ we denote
\beq\begin{split}\label{single_t}
& I_{r,J}(\bx_n;\bl_n;\bm{t}_{n-1};\epsilon)=e^{2\pi\iota x_n\underline{\boldsymbol{\l}}_n}\hat d_{n-1-r}\\[6pt]
& \times \int_{\mathbb{R}^{n-1-r}}d\bm{t}_{n-1-r}\; \hat\Psi_{(\bm{t}_{n-1-r} + (\iota g/2+\iota\epsilon)\,\bm{1}; \{\lambda_i+\iota g/2\}_{i\in J})}(\bx_{n-1}-x_n\bm{1})\\[6pt]
& \times \prod_{\substack{j\notin J\\l=1,\ldots,n-1-r}}\ccK(\lambda_j-t_l-\iota g/2-\iota\epsilon)\prod_{\substack{j\neq k\\1\leq j,k\leq n-1-r}}\ccm(t_j-t_k)\\[6pt]
& \times\prod_{\substack{j\in J\\ l=1,\ldots,n-1-r}}\ccK^{-1}(\lambda_j-t_l+\iota g/2-\iota\epsilon) \, \prod_{\substack{j \notin J\\k \in J}}\ccK(\l_j-\l_k-\iota g/2).
\end{split}\eeq
Passing from $\hat\Psi_{\bm{\l}_n(\bx_n)}$ to $\spsy\Phi_{\bm{\l}_n}(\bx_n)$, we multiply the integrand of (\ref{single_t}) by
\beq\begin{split}
F_1(\bx_n;&\bm{\l}_n;\bm{t}_{n-1-r};\epsilon) = e^{\pi g\sum_{1 \leq i < j \leq n}|x_i-x_j|} \prod_{1\leq i < j \leq n}\ccm(\lambda_i-\lambda_j) \\[6pt]
& \times e^{-\pi g\sum_{1 \leq i < j \leq n-1}|x_i-x_j|}\, \prod_{1\leq i < j \leq n-1-r}\ccm(t_i-t_j)^{-1}  \\[6pt]
& \times\prod_{\substack{i < j\\i,j\in J}}\ccm(\lambda_i-\lambda_j)^{-1} \, \prod_{\substack{i=1,\ldots,n-1-r\\j \in J}}\ccm(t_i-\lambda_j+\iota\epsilon)^{-1},
\end{split}\eeq
so (\ref{psi_dec}) becomes
\beq\begin{split}\label{phi_dec}
\spsy\Phi_{\bl_n}(\bx_n) = \sum^{n-1}_{r=0}\sum_{\substack{J \subset\{1,\ldots,n\}\\|J|=r}}\spsy I_{r,J}(\bx_n;\bl_n;\bm{t}_{n-1};\epsilon).
\end{split}\eeq
Denote additionally
\beqq\begin{split}
& F_2(\bx_n;\bm{\l}_n;\bm{t}_{n-1-r};\epsilon) = \spsy\Phi_{(\bm{t}_{n-1-r} + (\iota g/2+\iota\epsilon)\bm{1}; \{\lambda_i+\iota g/2\}_{i\in J})}(\bx_{n-1}-x_n\bm{1});\\[8pt]
& F_3(\bm{\l}_n;\bm{t}_{n-1-r};\epsilon) = \prod_{\substack{j\notin J\\l=1,\ldots,n-1-r}}\ccK(\lambda_j-t_l-\iota g/2-\iota\epsilon)\\[6pt]
& \times \prod_{\substack{j\neq k\\1\leq j,k\leq n-1-r}}\ccm(t_j-t_k) \prod_{\substack{j\in J\\ l=1,\ldots,n-1-r}}\ccK^{-1}(\lambda_j-t_l+\iota g/2-\iota\epsilon);\\[6pt]
& F_0(x_n;\bm{\l}_n) = e^{2\pi\iota x_n\underline{\boldsymbol{\l}}_n}\hat d_{n-1-r}; \qquad
F_4(\bm{\l}_n) = \prod_{\substack{j \notin J\\k \in J}}\ccK(\l_j-\l_k-\iota g/2).
\end{split}\eeqq
In these notations we have
\beqq\begin{split}
\spsy I_{r,J} = F_0\int_{\mathbb{R}^{n-1-r}}d\bm{t}_{n-1-r}\; F_2(\bm{t}_{n-1-r})F_3(\bm{t}_{n-1-r})F_4(\bl_n) F_1(\bm{t}_{n-1-r}).
\end{split}\eeqq
 By Stirling formula we have bounds 
\beq\begin{split}\label{kw_bnds}
& \big|\ccK(\lambda_j - t_l - \iota \alpha)\big| \leq C \, (1+|\Re\lambda_j-t_i|)^{g-1}e^{-\pi|\Re\lambda_j-t_l|},\\[6pt]
& \big|\ccm(t_i-t_j)\big| \leq C \, (1+|t_i-t_j|)^{1-g}e^{\pi|t_i-t_j|},\\[6pt]
& \big|\ccK^{-1}(\lambda_j-t_l + \iota\beta)\big| \leq C \, (1+|\Re\lambda_j-t_l|)^{1-g}e^{\pi|\Re\lambda_j-t_l|},
\end{split}\eeq
where $\alpha = g/2 + \epsilon$, $\beta = g/2-\epsilon$ and constants $C(g,\epsilon)$. Here we restrict $\epsilon > 0$ (kernel function has pole for $\epsilon = 0$).

Consider $J$ such that $r \leq n-2$. Bounds (\ref{kw_bnds}) imply the following estimate
\beq\begin{split}\label{gen_term}
& \Big|F_1(\bx_n;\bm{\l}_n;\bm{t}_{n-1-r};\epsilon)F_3(\bm{\l}_n;\bm{t}_{n-1-r};\epsilon)F_4(\bm{\l}_n)\Big| \leq C(g,\epsilon) \, e^{\pi g\sum\limits_{1\leq i\leq n-1}|x_i-x_n|}\\[6pt]
& \times \prod_{\substack{j < k\\ j,k\notin J}}(1+|\Re(\lambda_j-\lambda_k)|)^{1-g}  \prod_{\substack{j\notin J\\l=1,\ldots,n-1-r}}(1+|\Re\lambda_j-t_l|)^{g-1} \, \ddelta(\bm{t}_{n-1-r})^{1-g}\times\\[6pt]
&  \exp\pi\bigg\{\sum_{\substack{j<k\\j,k\notin J}}|\Re(\lambda_j-\lambda_k)| + \sum_{1\leq j<k\leq n-1-r}|t_j-t_k| -\sum_{\substack{j\notin J\\l=1,\ldots,n-1-r}}|\Re\lambda_j-t_l|\bigg\}.
\end{split}\eeq
It remains to bound $F_2$. According to shift relation $\spsy\Phi_{(\bm{\lambda}_n+\eta\bm{1})}(\bx_n) = e^{2\pi\iota\eta(x_1+\ldots x_n)} \spsy\Phi_{\bm{\lambda}_n}(\bx_n)$ and induction assumption, one can bound it as
\beq\begin{split}\label{gen_term2}
& \big|F_2(\bx_n;\bm{\l}_n;\bm{t}_{n-1-r};\epsilon)\big| \leq e^{-\pi g\sum_{1\leq i\leq n-1}(x_i-x_n)} \, \ddelta(\bm{t}_{n-1-r})^{N(n-1)}\\[6pt]
& \times\prod_{\substack{i<j\\i,j\in J}}(1+|\Re\lambda_i-\Re\lambda_j|)^{N(n-1)} \, \prod_{\substack{1\leq j\leq n-1-r\\i\in J}}(1+|\Re\lambda_i-t_j|)^{N(n-1)}\\
& \times e^{-2\pi\epsilon\sum_{1\leq i\leq n-1-r}(x_{r+i}-x_n)} \, \exp\bigg\{-2\pi\sum_{\substack{J=\{j_1<\ldots<j_r\}\\1\leq i\leq r}}(x_i-x_n)\Im\lambda_{j_i}\bigg\}.
\end{split}\eeq
Due to Proposition \ref{abc_int} combined with (\ref{gen_term}), (\ref{gen_term2}) and the following inequality
\beqq\begin{split}
e^{-2\pi\epsilon\sum_{1\leq i\leq n-1-r}(x_{r+i}-x_n)}  \exp\bigg\{-2\pi\sum_{j_1<\ldots<j_r} \sum_{1\leq i\leq r}(x_i-x_n)\Im\lambda_{j_i}\bigg\} \\[6pt]
\leq e^{-2\pi \sum_{1\leq i \leq n-1}(x_i-x_n)\Im\lambda_i},
\end{split}\eeqq
which holds due to assumption $\Im\lambda_1\leq \ldots\leq\Im\lambda_n <\epsilon$,
we arrive at
\beqq\begin{split}
\big|\spsy I_{r,J}\big| \leq C_{r,J}(g) \, \ddelta(\Re\bl_n)^{N(n)} \, e^{-2\pi(x_1\Im\l_1 +\ldots + x_n\Im\l_n)}.
\end{split}\eeqq
Consider the case $r = n-1$. By  (\ref{int_pres}) with $J = \{1,\ldots, n-1\}$
\beqq\begin{split}
\big|I_{n-1,\{1,\ldots,n-1\}}\big| = \Big|e^{2\pi\iota x_n(\lambda_1+\ldots+\lambda_n)}\hat\Psi_{(\lambda_1+\iota g/2,\ldots,\lambda_{n-1}+\iota g/2)}(\bx_{n-1}-x_n\bm{1})\\[6pt]
\times \prod_{1\leq i \leq n-1}\ccK(\lambda_n-\lambda_i - \iota g/2)\Big| \leq e^{-2\pi x_n\sum_i\Im\lambda_i}e^{-\pi g\sum_{1\leq i\leq n-1}(x_i-x_n)}\\[6pt]
\times \big|\hat\Psi_{(\lambda_1,\ldots,\lambda_{n-1})}(\bx_{n-1}-x_n\bm{1})\big| \, \prod_{1\leq i\leq n-1}\big|\ccK(\lambda_n-\lambda_i-\iota g/2)\big|.
\end{split}\eeqq
Passing from $\hat\Psi$ to $\spsy\Phi$, we multiply both sides by
\beqq\begin{split}
\frac{e^{\pi g\sum_{1 \leq i<j \leq n}|x_i-x_j|}\prod_{1 \leq i < j \leq n}\ccm(\lambda_i-\lambda_j)}{e^{\pi g\sum_{1\leq i<j\leq n-1}|x_i-x_j|}\prod_{1\leq i<j\leq n-1}\ccm(\lambda_i-\lambda_j)}.
\end{split}\eeqq
Due to $x_1 > \ldots > x_n$ and induction assumption, we obtain the final bound
\beqq\begin{split}
\big|\spsy I_{n-1,\{1,\ldots,n-1\}}\big| \leq C_{r,J}(g) \, \ddelta(\Re\bl_{n-1})^{N(n-1)} \, e^{-2\pi\sum_{1\leq i \leq n}x_i\Im\lambda_i}.
\end{split}\eeqq
Collecting all terms together, we have
\beqq\begin{split}
|\spsy\Phi_{\bl_n}(\bx_n)| \leq \sum_{J\subset\{1,\ldots,n\}}C_{|J|,J}(g) \,  \ddelta(\Re\bl_n)^{N(n)} \, e^{-2\pi\sum_{1\leq i \leq n}x_i\Im\lambda_i},
\end{split}\eeqq
which ends the proof.

To calculate $N(n)$, note that $N(2) = g$, thus we consider
\beqq\begin{split}
A = N(n-1), ~~~ B = N(n-1), ~~~ C = N(n-1)+1-g
\end{split}\eeqq
to obtain 
\beqq\begin{split}
N(n) = A + 2B + C + 1 = 4N(n-1) + 2 - g.
\end{split}\eeqq
It follows immediately that
\beqq\begin{split}
N(n) = 4^{n-2}g - \frac{(4^{n-2}-1)(g-2)}{3},
\end{split}\eeqq
as claimed in lemma.
\hfill{$\Box$} \bigskip

{\bf 2. Proof of Lemma \ref{rwf_comp}.} Consider the inequalities (\ref{gen_term}) and (\ref{gen_term2}) with zero imaginary parts $\Im\lambda_j = 0$ (note that these two formulas cover all terms except for asymptotical part, $r = n-1$). Specialization of (\ref{gen_term}) becomes
\beq\begin{split}\label{gen_t_re}
\big|F_1(\bx_n;\bm{\l}_n;\bm{t}_{n-1-r};\epsilon)F_3(\bm{\l}_n;\bm{t}_{n-1-r};\epsilon)F_4(\bm{\l}_n)\big| \leq C(g,\epsilon)\, e^{\pi g\sum\limits_{1\leq i\leq n-1}|x_i-x_n|} \,\\[6pt]
\times \prod_{\substack{j < k\\ j,k\notin J}}(1+|\lambda_j-\lambda_k|)^{1-g} \prod_{\substack{j\notin J\\l=1,\ldots,n-1-r}}(1+|\lambda_j-t_l|)^{g-1} \, \ddelta(\bm{t}_{n-1-r})^{1-g}\\
\times \exp\pi\bigg\{\sum_{\substack{j<k\\j,k\notin J}}|\lambda_j-\lambda_k| + \sum_{1\leq j<k\leq n-1-r}|t_j-t_k| - \sum_{\substack{j\notin J\\l=1,\ldots,n-1-r}}|\lambda_j-t_l|\bigg\}.
\end{split}\eeq
Analogously for (\ref{gen_term2}) 
\beq\begin{split}\label{gen_t_re2}
\bigg|F_2(\bx_n;\bm{\l}_n;\bm{t}_{n-1-r};\epsilon)\bigg| \leq e^{-\pi g\sum_{1\leq i\leq n-1}(x_i-x_n)} \, \ddelta(\bm{t}_{n-1-r})^{N(n-1)}\\[6pt]
\times\prod_{\substack{i<j\\i,j\in J}}(1+|\lambda_i-\lambda_j|)^{N(n-1)} \prod_{\substack{1\leq j\leq n-1-r\\i\in J}}(1+|\lambda_i-t_j|)^{N(n-1)}\\[6pt]
\times e^{-2\pi\epsilon\sum_{1\leq i\leq n-1-r}(x_{r+i}-x_n)}.
\end{split}\eeq
Applying Proposition \ref{abc_int} to the product of (\ref{gen_t_re}) and (\ref{gen_t_re2}), we have the bound
\beqq\begin{split}
\big|\spsy I_{r, J}\big| \leq C(g,\epsilon) \, \ddelta(\bl_n)^{N(n)} \, e^{-2\pi\epsilon\sum_{1\leq i\leq n-1-r}(x_{r+i}-x_n)}.
\end{split}\eeqq
Note that $|F_0(x_n;\bm{\l}_n)|\leq B(g)$. So, due to ordering $x_1 > \ldots > x_n$, summing over all $J$ with $|J| \leq n-2$, we arrive at the bound
\beqq\begin{split}
\big|\spsy\Phi_{\bx_n}(\bm{\lambda}_n) - \spsy\Phi^{as}_{\bx_n}(\bm{\lambda}_n)\big| \leq C(g, \epsilon) \, \ddelta(\bl_n)^{N(n)} \, e^{-2\pi\epsilon \, m_n(\bx_n)},
\end{split}\eeqq
where $m_n(\bx_n) =\min_i (x_i-x_{i+1})$. 
\hfill{$\Box$}
 
\subsection{Proof of Proposition \ref{abc_int}} \label{AppC}
Due to symmetry of the integrand, we can restrict the integration domain to the set
\beq\label{C1} X:=\{-\infty< t_1< t_2\ldots< t_{m-1}<\infty\}\eeq
and assume that the parameters $s_1,\ldots, s_m$ are also ordered,
\beq\label{C2} s_1< s_2<\ldots< s_m.\eeq
Let $\ppi$ be a linear order $\leq$ in the set $\{s_1,\ldots,s_m,t_1,\ldots t_{m-1}\}$ which is in accordance with \rf{C1} and \rf{C2}, that is $t_1\leq t_2\ldots\leq t_{m-1}$ and  $s_1\leq s_2\leq\ldots\leq s_m$. Each $\ppi$ determines a subdomain $X_\ppi\subset X$, where for each pair
$(t_k,s_l)$ the additional inequality $t_k< s_l$ or $s_l< t_k$ is imposed in accordance with the  linear order $\ppi$. We have a decomposition
\beq\label{C3} X=\bigsqcup_{\ppi}X_\ppi\eeq 
and prove the bound \rf{B108} for the integral over each $X_\ppi$.

From now on we fix parameters $s_i$ ordered as in \rf{C2} and a linear order $\ppi$. The integral \rf{B108} contains the exponent
$$\exp \pi\ Q(\bm{t}_{m-1};\bm{s}_{m})$$
where 
\beq\label{C4} Q(\bm{t}_{m-1};\bm{s}_{m})= \sum_{1\leq j<k\leq m}|s_j-s_k| + \sum_{1\leq j<k\leq m-1}|t_j-t_k| - \sum_{\substack{j=1,\ldots,m\\l=1,\ldots,m-1}}|s_j-t_l|.\eeq
Let us eliminate absolute values in \rf{C4}. Define intergers $\a(j)$ and $\b(j)$ by the relation
 \beq  \begin{array}{lll}\a(j)=2(k-j) \qquad \text{if} \qquad  t_j\in(s_k,s_{k+1}),\\[4pt]
 	\b(j)=2(k-j) \qquad\text{if} \qquad s_j\in(t_{k-1},t_{k}).\end{array}\eeq
 Then 
 \beq\label{C5} Q(\bm{t}_{m-1};\bm{s}_{m})=\sum_{j=1}^{m-1}\a(j)t_j+\sum_{k=1}^m\b(k)s_k.\eeq
 We call the variables $t_j$ with $\a(j)=0$ the variable of finite type, and the variables with
 $\a(j)\not=0$ the variables of infinite type. We further bound each integral over the variable of infinite type by  the integral over the ray going to plus  or minus infinity regarding to the sign of $\a(j)$, and then bound the integrals over variables $t_j$ of finite type by the integral over the finite segment $(s_j,s_j+1)$. 
 
  To do this, we use another combinatorial description of the linear order $\ppi$. Let us read the ordered
  set $\big(\{s_1,\ldots, s_m,t_1,\ldots t_{m-1}\},~\leq\big)$ from left to right, starting from its minimal element. Meeting the first neighboring pair $t_{i_1}<s_{j_1}$ or $s_{j_1}<t_{i_1}$ we set 
  $$\tau(i_1)=j_1$$
  then remove both variables $t_{i_1}$ and $s_{j_1}$ and repeat the procedure. We then get an inclusion $\tau$ of the set $\{1,\ldots ,m-1\}$ to the set  $\{1,\ldots ,m\}$. One can see that the formula \rf{C5} transforms to the expression
   \beq\label{C6} Q(\bm{t}_{m-1};\bm{s}_{m})=\sum_{j=1}^{m-1}\a(j)(t_j-s_{\tau(j)}).\eeq
 Let 
 $$t_{i_1},t_{i_2},\ldots,t_{i_n},\qquad i_1<i_2<\ldots< i_n$$ be all variables of finite type. Then the variables and parameters are organized in the following way:
 \beqq\begin{split} \{t_j,s_k|_{j,k<i_1}\},\ s_{i_1}, t_{i_1},&\  \{t_j,s_k|_{i_1<j,k<i_2}\}, \  s_{i_2}, t_{i_2},\  \{t_j,s_k|_{i_2<j,k<i_3}\}, \\ &\ldots\ ,  s_{i_n}, t_{i_n},\  \{t_j,s_k|_{i_n<j,k}\}. \end{split}\eeqq
Note that in each group $\{t_j,s_k|_{i_l<j,k<i_{l+1}}\}$:
\begin{enumerate}
	\item[(1)] all the variables $t_j$ are of infinite type;
	\item[(2)] the map $\tau$ sends each  set of indices $\{j:\,i_l<j<i_{l+1}\}$ to the set of indices
	$\{k:\,i_l<k<i_{l+1}\}$;
   \item[(3)] the sign of $\a(j)$ is the same for all $\{j:\,i_l<j<i_{l+1}\}$.
\end{enumerate}
We first bound the integral over the variables of infinite type.

Let $t_i$ be a variable of infinite type and $Y_i\subset\R$ be the region of its possible values inside the domaine $X_{\ppi}$, $u_1,\ldots,u_k\in\R$ and $p_1,\ldots p_k$ be non-negative real numbers, $p=p_1+\ldots +p_k$.  We claim that for some $C(p,\a(i))$, 
\beq \label{C7} \int_{Y_i}\prod_{k=1}^K (1+|t_i-u_k|)^{p_k}e^{\pi \a(i)(t_i-s_\tau(i))}dt_i<C(p,\a(i)) \prod_{k=1}^K(1+|s_{\tau(i)}-u_k|)^{p_k}.\eeq
Consider first the case $\a(i)<0$. Then $t_i>s_{\tau(i)}$, $Y_i\subset (s_{\tau(i)},+\infty)$, and thus
\beq \label{stepB} \begin{split}
		\int\limits_{Y_i}\prod_{k = 1}^K (1 + |t_i - u_k|)^{p_k}  e^{\pi\a(i)(t_i-s_\tau(i))}dt_i 
	\leq \int\limits^{+\infty}_{s_\tau(i)}\prod_{k = 1}^K (1 + |t_i - u_k|)^{p_k}  e^{\pi\a(i)(t_i-s_\tau(i))}dt_i 
	\\
	 \leq\int^{+\infty}_{s_{\tau(i)}}\prod^K_{k=1}(1+|t_i-s_{\tau(i)}|)^{p_k}(1+|s_{\tau(i)}-u_k|)^{p_k} \, e^{\pi\a(i)(t_i-s_{\tau(i)})}dt_i \\
	= \prod^K_{k = 1}(1+|s_{\tau(i)}-u_k|)^{p_k}\int^{+\infty}_{s_1}(1+(t_i-s_\tau(i)))^{p} e^{\pi\a(i)(t_i-s_\tau(i))}dt_i\\
	= \prod^K_{k = 1}(1+|s_{\tau(i)}-u_k|)^{p_k} \int^{+\infty}_0(1+t)^{p}e^{\pi\a(i) t}dt \\
	 = \prod^K_{k = 1}(1+|s_{\tau(i)}-u_k|)^{p_k}e^{-\pi\a(i)}(-\pi\a(i))^{-1-p} \int^{+\infty}_{-\pi\a(i)}t^{p}e^{- t}dt\\
	\leq \frac{\Gamma(p - 1)e^{-\pi\a(i)}}{(-\pi\a(i))^{p + 1}}\prod^K_{k = 1}(1+|s_1-u_k|)^{p_k}.
\end{split}\eeq
The case $\a(i)>0$ is analogous: we analyse the integral over $(-\infty,s_{\tau(i)})$ and get  the same bound.

The remaining integrals over the variables $t_i$ of finite type we bound just by the maximum of the integrand on the segment $[s_i,s_{i+1}]$ multiplied by the length $s_{i+1}-s_i$ of the segment:
\beqq \begin{split} 
	& \int_{Y_i}\prod_{k=1}^K (1+|t_i-s_k|)^{p_k}dt_i\leq  \int_{s_i}^{s_{i+1}}\prod_{k=1}^K (1+|t_i-s_k|)^{p_k}dt_i \\[6pt]
	& \leq (s_{i+1}-s_i) \,	\prod_{s_k\leq s_i}(1+|s_{i+1}-s_k|)^{p_k} \,	\prod_{s_k\geq s_{i+1}}(1+|s_{i}-s_k|)^{p_k}.
	\end{split}\eeqq
Let us summarize the above calculations. We bound part of the integral \rf{B108} over subdomain $X_{\ppi}$ by the product
\beq\label{C8} D \, \prod_{1\leq a<b\leq m}(1+|s_a-s_b|)^{k_{ab}B +l_{ab}C} \, \prod_{k=1}^n(s_{i_{k+1}}-s_{i_k}),\eeq
where the constant $D$ depends on $B,C$ and $\ppi$, the nonnegative integers $k_{ab}$ do not exceed~$2$, nonnegative integers~$l_{ab}$ do not exceed $1$ except for the cases when 
 \beq\label{C9} b=i_k+1\qquad \text{for some} \qquad k=1,\ldots,n \eeq
where $n$ is the number of the variables of the finite type.
Here 
\beq\label{C10} k_{a,i_k+1}=3, \qquad l_{a,i_k+1}=2. \eeq
On the other hand, we have
\beq\label{C11} k_{a,i_k}=1, \qquad l_{a,i_k}=0. \eeq
We then can replace the factors 
\beq\label{C12}(1+|s_a-s_{i_k+1}|)^B \qquad \text{and}\qquad (1+|s_a-s_{i_k+1}|)^C\eeq
by the bigger ones
\beq\label{C13}(1+|s_a-s_{i_{k+1}}|)^B \qquad \text{and}\qquad (1+|s_a-s_{i_{k+1}}|)^C\eeq
if $k<n$ and by 
\beq\label{C13-2}(1+|s_a-s_{\theta}|)^B \qquad \text{and}\qquad (1+|s_a-s_{\theta}|)^C\eeq
if $k=n$. Here $\theta$ is the only element of the set $\{1,\ldots, m\}$ which is not in the image of the map $\tau$. After that replacement we arrive at the bound
\beq\label{C14}\begin{split}& D \cdot \prod_{1\leq a<b\leq m}(1+|s_a-s_b|)^{A+2B +C} \, \prod_{k=1}^n(s_{i_{k+1}}-s_{i_k})\leq \\ &D\cdot  \prod_{1\leq a<b\leq m}(1+|s_a-s_b|)^{A+2B+C+1},\end{split}\eeq
as stated in proposition.
\hfill{$\Box$}

\section{Evaluation at zero point}
\label{Norm}

Let us calculate the value of the eigenfunction 
$\Psi_{\bm{\lambda}_n}(\bm{x}_n)$ at special point 
$x_1=\ldots=x_n=0$ using iterative structure \eqref{g11} of 
Mellin--Barnes representation \eqref{Psi-MB}.
We have 
\begin{align*}
\Psi_{\bm{\lambda}_n}(\bm{x}_n)= \hat{\Lambda}_n(x_n)\,
\Psi_{\bm{\lambda}_{n-1}}(\bm{x}_{n-1})
\end{align*}
or in explicit form
\begin{align*}
\Psi_{\bm{\lambda}_n}(\bm{x}_n) &= 
\frac{\Gamma^{1-n}(g)}{(n-1)!} \, \int_{\mathbb{R}^{n - 1}}\,
\exp \biggl( 2\pi \imath x_n \biggl( \sum_{i = 1}^{n} 
\lambda_{i} - \sum_{i = 1}^{n - 1} \gamma_{i} \biggr) \biggr) \\[6pt]  
& \times \frac{ \prod_{i = 1}^{n} \prod_{j = 1}^{n-1} \, \Gamma \bigl( \imath \lambda_{i} - \imath  \gamma_{j} + \frac{g}{2} \bigr) \, \Gamma \bigl( \imath \gamma_{j} - \imath  \lambda_{i} + \frac{g}{2} \bigr) }{\prod_{1 \leq i \not= j \leq n-1} \, \Gamma(\imath \gamma_{i} - \imath \gamma_{j}) \Gamma(\imath \gamma_{j} - \imath \gamma_{i} + g)} \, \Psi_{\bm{\gamma}_{n-1}}(\bm{x}_{n-1})
\prod_{k = 1}^{n - 1} \frac{d\gamma_{k}}{2\pi}.
\end{align*}
After substitution $x_1=\ldots=x_n=0$ we obtain the useful iterative formula 
\begin{align*}
\Psi_{\bm{\lambda}_n}(0) = 
\frac{\Gamma^{1-n}(g)}{(n-1)!} \, \int_{\mathbb{R}^{n - 1}}\,
\frac{ \prod_{i = 1}^{n} \prod_{j = 1}^{n-1} \, \Gamma \bigl( \imath \lambda_{i} - \imath  \gamma_{j} + \frac{g}{2} \bigr) \, \Gamma \bigl( \imath \gamma_{j} - \imath  \lambda_{i} + \frac{g}{2} \bigr) }{\prod_{1 \leq i \not= j \leq n-1} \, \Gamma(\imath \gamma_{i} - \imath \gamma_{j}) \Gamma(\imath \gamma_{j} - \imath \gamma_{i} + g)} \, \Psi_{\bm{\gamma}_{n-1}}(0)
\prod_{k = 1}^{n - 1} \frac{d\gamma_{k}}{2\pi}.
\end{align*}
Let us illustrate the main idea of calculation using the simplest examples. We start with the case $n=2$ and use the initial 
condition $\Psi_{\lambda_1}(x_1)=e^{\imath \lambda_1 x_1}$  
so that $\Psi_{\lambda_1}(0)=1$. 
We have 
\begin{align*}
\Psi_{\lambda_1\lambda_2}(0) = 
\frac{1}{\Gamma(g)} \, \int_{\mathbb{R}}
\prod_{k = 1}^{2}
\Gamma \bigl( \imath \lambda_k + {\textstyle \frac{g}{2}} - \imath  \gamma \bigr) 
\Gamma \bigl( \imath \gamma - \imath \lambda_k + {\textstyle \frac{g}{2}} \bigr)
\frac{d\gamma}{2\pi} = \frac{\Gamma(g)}{\Gamma(2g)}\,
\Gamma(g+\imath \lambda_{12})\Gamma(g+\imath \lambda_{21}).
\end{align*}
The integral is calculated by using the Barnes lemma \cite[Theorem 2.4.2]{AAR}
\begin{align}
\int_{\mathbb{R}} 
\prod_{k = 1}^{2}
\Gamma \bigl( \alpha_k - \imath  \gamma \bigr) 
\Gamma \bigl( \imath \gamma - \beta_k \bigr)
\frac{d\gamma}{2\pi} = 
\frac{\prod_{i,k=1}^2\,\Gamma(\alpha_i-\beta_k)}
{\Gamma(\alpha_1+\alpha_2-\beta_1-\beta_2)}
\end{align}
for $\alpha_k = \imath \lambda_k + {\textstyle \frac{g}{2}}$ and 
$\beta_k = \imath \lambda_k - {\textstyle \frac{g}{2}}$.
Integration is performed along real axis, which separates infinite series of poles 
$\gamma = \lambda_k-\imath{\textstyle \frac{g}{2}}-\imath n\,,n=0,1,2,\ldots$ in lower half-plane arising 
from $\Gamma(\alpha_k-i\gamma)$ and infinite series of poles 
$\gamma = \lambda_k+\imath{\textstyle \frac{g}{2}}+\imath n\,,n=0,1,2,\ldots$
in upper half-plane arising from $\Gamma(i\gamma-\beta_k)$,
so that the required condition of Barnes lemma is satisfied.

The next nontrivial example is $n=3$. For brevity, denote $\gamma_{ij} = \gamma_i - \gamma_j$, then we have
\begin{multline*}
\Psi_{\lambda_1\lambda_2\lambda_3}(0) = 
\frac{\Gamma^{-2}(g)}{2!} \, \int_{\mathbb{R}^2}\,
\frac{ \prod\limits_{i = 1}^{3} \prod\limits_{j = 1}^{2} \, \Gamma \bigl( \imath \lambda_{i} - \imath  \gamma_{j} + \frac{g}{2} \bigr) \, \Gamma \bigl( \imath \gamma_{j} - \imath  \lambda_{i} + \frac{g}{2} \bigr) }{\Gamma(\imath \gamma_{12})
\Gamma(\imath \gamma_{21}) 
\Gamma(\imath \gamma_{12} + g)
\Gamma(\imath \gamma_{21} + g)} \, \Psi_{\gamma_1\gamma_2}(0)
\prod\limits_{k = 1}^{2} \frac{d\gamma_{k}}{2\pi} \\[6pt] 
= \frac{1}{2\Gamma(g)\Gamma(2g)}\, 
\int\limits_{\mathbb{R}^2}\,
\frac{ \prod\limits_{i = 1}^{3} \prod\limits_{j = 1}^{2} \, \Gamma \bigl( \imath \lambda_{i} - \imath  \gamma_{j} + \frac{g}{2} \bigr) \, \Gamma \bigl( \imath \gamma_{j} - \imath  \lambda_{i} + \frac{g}{2} \bigr) }{\Gamma(\imath \gamma_{12})
\Gamma(\imath \gamma_{21})} \, 
\prod_{k = 1}^{2} \frac{d\gamma_{k}}{2\pi} \\[6pt] 
= \frac{\Gamma(g)}{\Gamma(2g)}\,\frac{\Gamma(g)}{\Gamma(3g)}\,
\prod_{1 \leq i \not= j \leq 3} \, \Gamma(\imath \lambda_{ij} + g).
\end{multline*}
In the last line we used Gustafson integral of A-type \cite[Theorem 5.1]{G}, which generalizes the Barnes lemma,
\begin{align}\label{gus-1}
\int_{\mathbb R^n}\,
\frac{\prod\limits_{k=1}^{n+1}\prod\limits_{j=1}^n \Gamma( \alpha_k- \imath \lambda_j)\Gamma(\imath\lambda_j-\beta_k)}
{\prod_{j < k}\Gamma(\imath\lambda_k-\imath\lambda_j)
\Gamma(\imath\lambda_j-\imath\lambda_k)}
\prod_{j=1}^n \frac{d\lambda_j}{2\pi} =\frac{n!\,\prod_{k,j=1}^{n+1}\Gamma(\alpha_j-\beta_k)}
{\Gamma\left(\sum_{j=1}^{n+1}(\alpha_j-\beta_j)\right)}.
\end{align}
Integration is performed along real axis in each variable $\lambda_j $ and it is assumed that real axis separates infinite series of poles in lower half-plane arising from $\Gamma(\alpha_k-i\lambda_j)$ and infinite series of poles
in upper half-plane arising from $\Gamma(i\lambda_j-\beta_k)$.
We used \eqref{gus-1} for $n=2$ and $\alpha_k = \imath \lambda_k + {\textstyle \frac{g}{2}}$ and $\beta_k = \imath \lambda_k - {\textstyle \frac{g}{2}}$, so that the required condition is satisfied.

We hope that the whole mechanism is clear and 
the form of the general expression  
\begin{align}\label{Psi0}
\Psi_{\bm{\lambda}_n}(0) = \prod_{k=1}^n \frac{\Gamma(g)}{\Gamma(kg)}\,
\prod_{1 \leq i \not= j \leq n} \, \Gamma(\imath \lambda_{ij} + g)
\end{align}
equivalent to \rf{norm} looks naturally. It remains to prove it using induction and of course everything is based on the Gustafson integral \eqref{gus-1}.
We have 
\begin{multline*}
\Psi_{\bm{\lambda}_{n+1}}(0) = 
\frac{\Gamma^{-n}(g)}{n!} \, \int\limits_{\mathbb{R}^{n}}\,
\frac{ \prod_{i = 1}^{n+1} \prod_{j = 1}^{n} \, 
\Gamma \bigl( \imath \lambda_{i} - \imath  \gamma_{j} + \frac{g}{2} \bigr) \, \Gamma \bigl( \imath \gamma_{j} - \imath  \lambda_{i} + \frac{g}{2} \bigr) }{\prod_{1 \leq i \not= j \leq n} \, \Gamma(\imath \gamma_{i} - \imath \gamma_{j}) \Gamma(\imath \gamma_{j} - \imath \gamma_{i} + g)} \, \Psi_{\bm{\gamma}_{n}}(0)
\prod_{k = 1}^{n} \frac{d\gamma_{k}}{2\pi}
 \\[6pt] 
= \Gamma^{-n}(g)\,
\prod_{k=1}^n \frac{\Gamma(g)}{\Gamma(kg)}\,
\frac{1}{n!} \, \int_{\mathbb{R}^{n}}\,
\frac{ \prod\limits_{i = 1}^{n+1} \prod\limits_{j = 1}^{n} \, 
\Gamma \bigl( \imath \lambda_{i} - \imath  \gamma_{j} + \frac{g}{2} \bigr) \, \Gamma \bigl( \imath \gamma_{j} - \imath  \lambda_{i} + \frac{g}{2} \bigr) }{\prod_{1 \leq i \not= j \leq n} \,\Gamma(\imath \gamma_{i} - \imath \gamma_{j})  }\,
\, \prod_{k = 1}^{n} \frac{d\gamma_{k}}{2\pi}   \\[6pt]
= \Gamma^{-n}(g)\,
\prod_{k=1}^n \frac{\Gamma(g)}{\Gamma(kg)}\,
\frac{\Gamma^{n + 1}(g)\,\prod_{1 \leq i \not= j \leq n+1} \, \Gamma(\imath \lambda_{ij} + g)}{\Gamma((n+1)g)} = \prod_{k=1}^{n+1} \frac{\Gamma(g)}{\Gamma(kg)}\,
\prod_{1 \leq i \not= j \leq n+1} \, \Gamma(\imath \lambda_{ij} + g),
\end{multline*}
where we used \eqref{gus-1} for $\alpha_k = \imath \lambda_k + {\textstyle \frac{g}{2}}$ and $\beta_k = \imath \lambda_k - {\textstyle \frac{g}{2}}$.

		\end{document}